\documentclass[times,twocolumn,final]{elsarticle}

\usepackage{float}

\usepackage{medima}
\usepackage{framed,multirow}

\usepackage{amssymb}
\usepackage{latexsym}
\usepackage{bbold}

\usepackage{url}
\usepackage{xcolor}

\usepackage{hyperref}

\usepackage{amsthm,amssymb}
\usepackage{amsmath,etoolbox}
\usepackage{mathtools}
\usepackage{filecontents,catchfile}
\usepackage[ruled]{algorithm2e}
\usepackage{graphicx,subfigure}

\definecolor{newcolor}{rgb}{.8,.349,.1}

\journal{Medical Image Analysis}

\begin{document}

\verso{Yiwen Li \textit{et~al.}}

\begin{frontmatter}

\title{Prototypical few-shot segmentation for cross-institution male pelvic structures with spatial registration}%


\author[1]{Yiwen \snm{Li}\corref{cor1}}
\cortext[cor1]{Corresponding author: 
  E-mail: yiwen.li@st-annes.ox.ac.uk}
\author[2,3]{Yunguan \snm{Fu}}
\author[2]{Iani J.M.B. \snm{Gayo}}
\author[2]{Qianye \snm{Yang}}
\author[2]{Zhe \snm{Min}}
\author[2]{Shaheer U. \snm{Saeed}}
\author[2,7]{Wen \snm{Yan}}
\author[4]{Yipei \snm{Wang}} 
\author[4]{J. Alison \snm{Noble}}
\author[5]{Mark \snm{Emberton}}
\author[2]{Matthew J. \snm{Clarkson}}
\author[6]{Henkjan \snm{Huisman}}
\author[2]{Dean C. \snm{Barratt}}
\author[1]{Victor A. \snm{Prisacariu}}
\author[2,4]{Yipeng \snm{Hu}}

\address[1]{Active Vision Laboratory, Department of Engineering Science, University of Oxford, Oxford, U.K.}
\address[2]{Department of Medical Physics and Biomedical Engineering, UCL Centre for Medical Image Computing, and Wellcome / EPSRC Centre for Interventional and Surgical Sciences, University College London, London, U.K.}
\address[3]{InstaDeep Ltd., London, U.K.}
\address[4]{Institute of Biomedical Engineering, Department of Engineering Science, University of Oxford, Oxford, UK}
\address[5]{Division of Surgery \& Interventional Science, University College London, London, U.K.}
\address[6]{Department of Radiology, Radboud University Nijmegen Medical Centre, Nijmegen, The Netherlands}
\address[7]{Department of Electrical Engineering, City University of Hong Kong, Hong Kong, China}


\begin{abstract}
The prowess that makes few-shot learning desirable in medical image analysis is the efficient use of the \textit{support} image data, which are labelled to classify or segment new classes, a task that otherwise requires substantially more training images and expert annotations. 
This work describes a fully 3D prototypical few-shot segmentation algorithm, such that the trained networks can be effectively adapted to clinically interesting structures that are absent in training, using only a few labelled images from a different institute. First, to compensate for the widely recognised spatial variability between institutions in episodic adaptation of novel classes, a novel \textit{spatial registration mechanism} is integrated into prototypical learning, consisting of a segmentation head and an spatial alignment module. Second, to assist the training with observed imperfect alignment, \textit{support mask conditioning module} is proposed to further utilise the annotation available from the support images. 
Extensive experiments are presented in an application of segmenting eight anatomical structures important for interventional planning, using a data set of 589 pelvic T2-weighted MR images, acquired at seven institutes. The results demonstrate the efficacy in each of the 3D formulation, the spatial registration, and the support mask conditioning, all of which made positive contributions independently or collectively. Compared with the previously proposed 2D alternatives, the few-shot segmentation performance was improved with statistical significance, regardless whether the support data come from the same or different institutes. 

\end{abstract}


\end{frontmatter}



\newcommand{\codeurl}{\url{https://github.com/kate-sann5100/CrossInstitutionFewShotSegmentation}}
\newcommand{\dataurl}{\url{https://zenodo.org/record/7013610}}

\section{Introduction}\label{sec:intro}

Multi-structure segmentation is one of the fundamental computing tasks in medical imaging applications, found in diagnosis, treatment, and monitoring, and remains a research interest.
Diagnosis of varieties of diseases can be assisted by quantifying the morphology, or its change, of multiple structures. For example, brain disorders, including Alzheimer’s disease~\citep{petrella2003neuroimaging} and Parkinson~\citep{hutchinson2000structural}, associates with abnormal volumes or shapes of neurological regions. Identifying brain structures is the key to many quantitative studies, such as functional activation mapping and brain development analysis~\citep{han2007atlas}.
Minimally invasive treatments often benefit from careful planning of both interventional instruments and guidance imaging, with respect to segmented patient-specific anatomical structures. In endoscopic pancreatobiliary procedures, as previously reported, image guidance that displays registered anatomical models outside the endoscopic field of view helps the surgeon during targeting and navigation~\citep{howe1999robotics}.

This work is primarily concerned with segmenting multiple organs and urologically interesting structures on T2-weighted MR images from prostate cancer patients, to plan targeted biopsy, focal therapy, and, increasingly, other therapeutic procedures such as radiotherapy. Accurate segmentation of these structures help with targeting suspected cancerous regions found in multiparametric MR imaging with respect to the prostate gland and avoiding vulnerable surrounding structures, such as rectum, bladder and neurovascular bundles, for minimising risks in infection, impotence and other potential injury and complications~\citep{de2010focal}.

The data-driven representation learning enabled by deep neural networks has led to promising segmentation results in multi-structure segmentation tasks, for example, in neuroimaging~\citep{henschel2020fastsurfer} and abdominal organs ~\citep{weston2019automated}. Automating this task reduces the current requirement of manual segmentation which is often associated with costs in expertise and intra- and inter-observer variations~\citep{fiorino1998intra}. However, recent supervised segmentation methods mostly rely on large data sets with full annotations, subject to the similar limitations in labelling, albeit only with training. On the other hand, few-shot image learning aims to classify unseen classes using only a few labelled examples~\citep{snell2017prototypical, sung2018learning}. For medical image segmentation, such novel classes may represent new types of organ or anatomical regions, whose expert annotations are not available for large training data sets. Using our intended interventional planning application as an example, different biopsy or therapy procedures may require different anatomical structures and pathological regions to be annotated during the planning stage. Prostate zonal structures, if not routinely segmented for MR-planning of radiotherapy, may provide more precise target localisation in registration-assisted ultrasound-guided focal ablation or a new therapy.

\begin{figure}[!t]
\centering
\includegraphics[scale=.3]{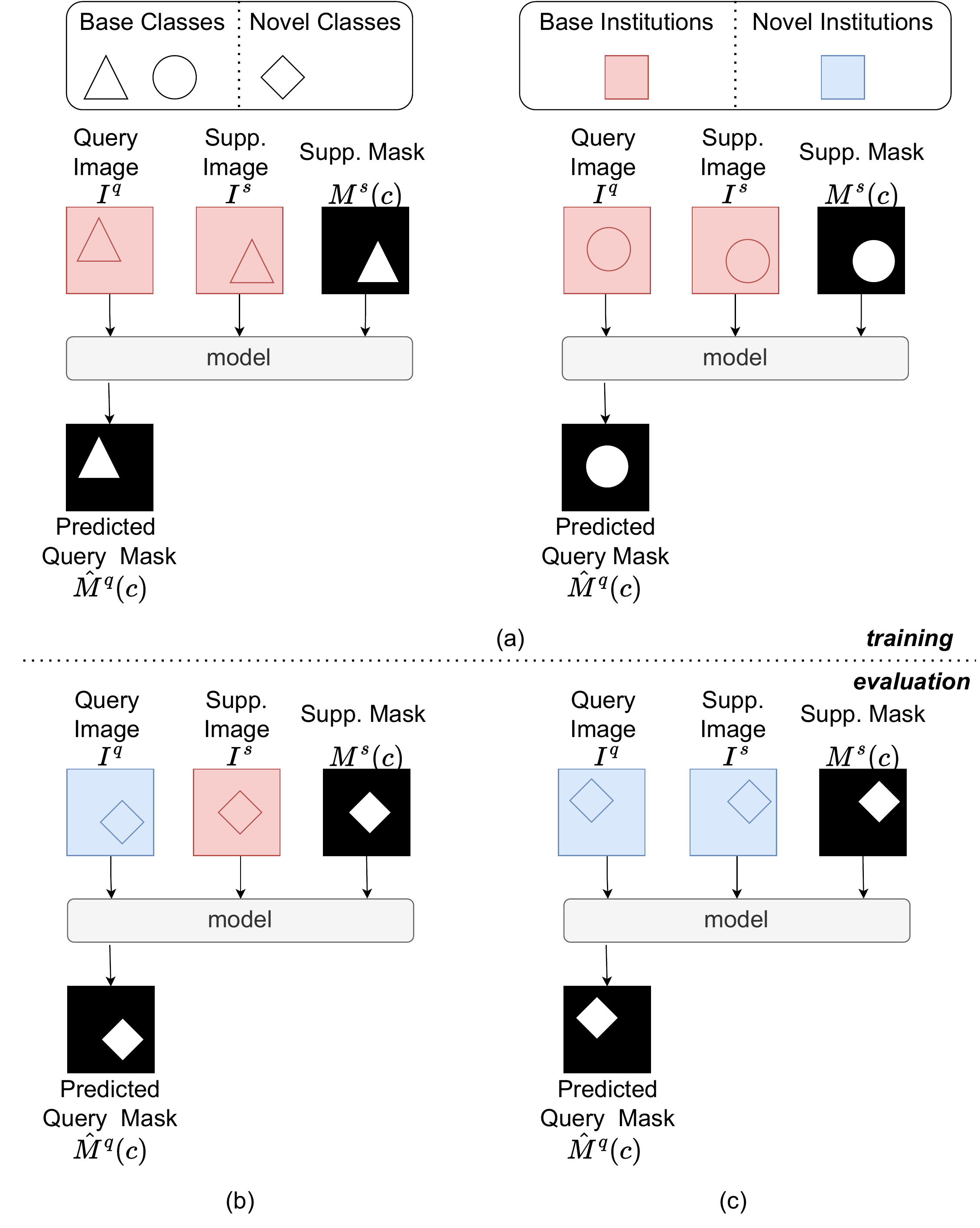}
\caption{
Visualisation of the proposed cross-institution few-shot segmentation task. As shown in (a), the task aims to train a model on the base data set including images from base institutions with corresponding masks on base classes and generalise to novel institution as well as novel classes. The model is evaluated on segmenting novel classes from query images acquired by novel institutions, while support images comes from either base or novel institutions, respectively shown in (b) and (c).}
\label{fig:pipeline}
\end{figure}

The now well-recognised performance loss in cross-institute generalisation of deep learning models \citep{gibson2018inter} has motivated a body of research such as domain adaptation~\citep{ren2018adversarial, meyer2021domain} and federated learning~\citep{li2021fedbn}. This work focuses on few-shot segmentation with the practically important cross-institution context (as shown in Fig.~\ref{fig:pipeline}), which aims to segment a \emph{novel class} from a \emph{query} image, given only a few \emph{support} images and their binary masks of the novel class, from a \textit{novel institution} where the limited labelled data are available. 
In other words, the model should be able to simultaneously adapt to both novel classes and novel institutions.

To improve the inter-class and inter-institute generalisation, this work first examines the manifestation of the performance-reducing inter-institute variability in prototypical few-shot image segmentation algorithms. In particular, we investigate the impact of spatial alignment, or the lack thereof, between support and query data from different institutes, a key component in such prototypical learning paradigm, on the few-shot segmentation accuracy.

First, addressing one of the previously identified challenges of spatial inconsistency~\citep{tian2020prior}, also found in medical image applications~\citep{guo2021multi,sun2022few}, we develop a \textit{spatial registration mechanism} to align the support and query images prior to the comparison between the two. This spatial registration mechanism consists of a \textit{segmentation head} and a \textit{spatial alignment module}, trained end-to-end, and is motivated by medical-image-specific observations of the difference between intra- and inter-institution data characteristics, due to different scanners and local imaging protocols, discussed further in Section~\ref{sec:align}. 

Second, we propose an additional \textit{support mask conditioning module}, also trained end-to-end, to enforce the conditioning on the available novel class labels. The conditioning module is empirically designed to work together with the spatial registration mechanism, to maximise utilisation of the few and prized support masks.

In addition to the evaluation of efficacy due to each proposed component, this work also demonstrates the benefit of the proposed 3D formulation, replacing existing 2D neural network-based few-shot segmentation approaches~\citep{feyjie2020semi,ouyang2020self,roy2020squeeze,abdel2021fss,guo2021multi,tang2021recurrent,yu2021location,sun2022few}.  

One widely identified problem in both developing and evaluating multi-structure segmentation algorithms is the lack of a sizable labelled data set. To the date of submission of this paper, there were no multi-structure annotations publicly available for pelvic MR images. Through this work, all of our manual labels from open data sets have been made available at \dataurl, for aiding the reproducibility of this work and, potentially, for other urologic or radiologic tasks concerning multiple pelvic anatomical structures.

Our preliminary results were recently presented~\citep{li2022few}, contributions from this paper include 1) a more detailed description and discussion of the spatial registration mechanism, 2) a new support mask conditioning module, 3) a substantially larger and fully labelled data set, and 4) more ablation comparison experiments. These are also summarised as follows.

\begin{itemize}
    \item We introduced the cross-institution few-shot segmentation task to address the data scarcity problem specifically faced in medical applications.
    \item We first proposed to extend prototypical neural network to 3D for few-shot multi-class segmentation, which requires fewer parameters while achieving similar performance to its 2D counterpart.
    \item We developed an spatial registration mechanism and a support mask conditioning module, directly addressing the observed limitations in medical image few-shot segmentation, for improving generalisation across data from different institutions.
    \item We presented extensive ablation studies to investigate the impact of the proposed individual components, the increasing number of support data, the varying size of the training set, and the permutations in the available institutes.
    \item \textcolor{black}{We published all expert annotations based on public image data sets at \dataurl, which includes full segmentation of eight distinctive male lower pelvic structures on 589 3D MR images (including 178 3D MR images from our preliminary work~\citep{li2022few}).}
    \item The code implementing the proposed algorithms has also been made publicly available at \codeurl.
\end{itemize}

\section{Related Work}
\label{sec:related_work}


\subsection{Few-shot segmentation}
The few-shot segmentation task was first introduced in computer vision applications~\citep{shaban2017one} where the goal is to segment the \emph{novel class} in a \emph{query} image in the presence of a few \emph{support} images having the same class labelled. Using episodic training which takes both query and support images as input, \cite{shaban2017one} demonstrated a better performance compared to the common fine-tuning methods, which fine-tunes the models on the support images per novel class. In 2018, \citet{dong2018few} proposed prototypical episode learning that represents the novel class in the support image with a single \emph{prototype} vector and compares it with query features to perform segmentation. This strategy was later adopted in many further research works \citep{zhang2019pyramid,liu2020part,li2021few}.

Due to the common challenges in data collection, few-shot segmentation was also adapted to different medical images, including CT~\citep{roy2020squeeze}, MRI~\citep{mondal2018few}, ultrasound ~\citep{guo2021multi}, etc. The early methods applied fine-tuning strategy and addressed the over-fitting on support images with multi-tasking \citep{mondal2018few,cui2020unified} and data augmentation \citep{zhao2019data,he2020deep,wang2021few}.
\citet{roy2020squeeze} was one of the first that adopted prototypical learning in medical imaging, reporting promising performance. \cite{ouyang2020self} and \cite{yu2021location} extracted multiple prototype vectors and performed a location-guided comparison, with the assumption of similar spatial layouts between the query and support images. However, in cross-institution scenarios, regions of interest may be located differently between queries and supports, as shown in Fig.~\ref{fig:align_vis}. In this work, we proposed an integrated spatial registration mechanism to address such inconsistency. 

In addition to data scarcity, higher-dimensional data in medical imaging often poses practical challenges in neural network training. \citet{roy2020squeeze} proposed to use 2D neural networks, that pre-trained on large data sets such as ImageNet, and performed slice-by-slice inference when applied to 3D medical images. 
This strategy was also adopted by most of the follow-up prototypical methods~\citep{feyjie2020semi,ouyang2020self,abdel2021fss,guo2021multi,tang2021recurrent,yu2021location,sun2022few}. \citet{kim2021bidirectional} integrated a bidirectional gated recurrent unit to process 2D features extracted from adjacent slices. \citet{zhou2021generalized} proposed 3D pyramid reasoning modules (PRMs) to model the anatomical correlation between query features at each location and all support features at neighbouring corresponding locations. To reduce computational cost, a relatively small number of channels were used for each convolutional kernel. The proposed method, in contrast, reduced the number of comparisons by extracting a single prototype vector for each spatial window.

To the best of our knowledge, there has been no prior work that successfully deployed 3D neural networks that receive and output 3D image volumes directly, using prototypical training for few-shot segmentation in medical imaging applications. Investigating the 3D formulation is not only technically interesting, but may also lead to potentially superior performance and/or efficiency in this inherently 3D segmentation task.

\subsection{Cross-institution learning}
The proposed cross-institution few-shot segmentation task aims to segment a novel class in images from a novel institution, with support images and labels from the same novel or other non-novel institutions. Although the objective is also to generalise on novel data, it differs from federated learning and domain adaptation due to their constraints on data privacy, accessibility, and availability. However, an optimal gain in efficient use of labelled data could be achieved by combining these methods with few-shot learning.

Federated learning is a learning paradigm that targets the problem of data governance and privacy by training algorithms collaboratively without the need for physically exchanging the data themselves, sometimes requiring the compliance of varying access policy~\citep{rieke2020future}. It has been used in different medical imaging applications~\citep{li2021fedbn}, but the focus is mainly on improving performance in trained classes without the need for generalisation in a novel class.

There exist three different domain adaptation methods: supervised, unsupervised, and semi-supervised. While sharing the same overarching objective, i.e. generalising to new domains such as novel classes, the supervised domain adaptation~\citep{hosseini2016alzheimer} does not explicitly focus on the data scarcity as few-shot learning. Unsupervised methods~\citep{perone2019unsupervised}, on the other hand, often assume large-scale data set from the novel institution but without labels, thus paying a different attention from that of cross-institution few-shot learning. Combining both techniques, semi-supervised methods~\citep{fu2019more,xia2020uncertainty} can leverage labelled and unlabelled data sets more efficiently.

Despite the distinct focuses, techniques and methodologies from federated learning and domain adaptation have indeed been considered in developing our cross-institute few-shot segmentation approach. For example, the spatial normalisation and divergence between features have commonly been adopted in federated learning~\citep{li2020predicting} and feature-level domain adaptation~\citep{tomar2021self}, respectively.



\begin{figure*}[h!]
\centering
\includegraphics[scale=.5]{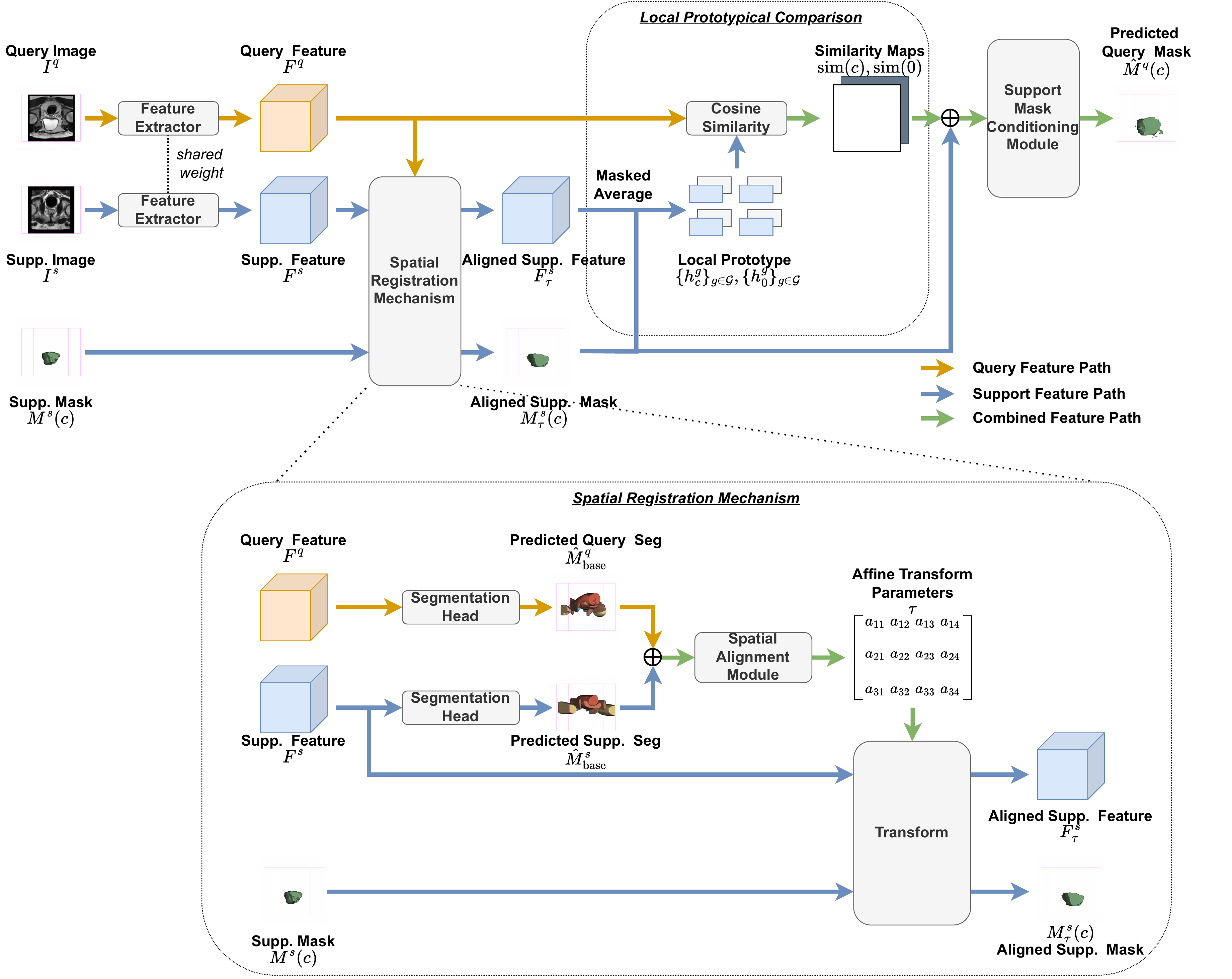}
\caption{Overview of the proposed method. A shared feature extractor outputs query/support features from query/support images. The spatial registration mechanism spatially registers support feature/mask towards query. Foreground/background similarity maps are derived through local prototypical comparison, concatenated with the aligned support mask and processed by the support mask conditioning module to make the final prediction}
\label{fig:3d_con_align}
\end{figure*}

\section{The Cross-Institution Few-Shot Segmentation Task}
\label{sec:task}

Consider a set of classes and institutions, $\mathcal{C}$ and $\mathcal{U}$, respectively. For each institution $u\in\mathcal{U}$, $\mathcal{I}_u$ denotes the set of all images. All classes have been segmented for each image: given a class $c\in\mathcal{C}$, $M(I_u, c)$ represents the corresponding mask for the image $I_u\in\mathcal{I}_u$ from the institution $u$.

The classes $\mathcal{C}$ and the institutions $\mathcal{U}$ are split into disjoint sets $(\mathcal{C}_\text{base}, \mathcal{C}_\text{novel})$ and $(\mathcal{U}_\text{base}, \mathcal{U}_\text{novel})$, respectively. The images $\mathcal{I}_u$ of each institution $u$ are also separated into disjoint training and test subsets $\mathcal{I}^\text{train}_u$ and $\mathcal{I}^\text{test}_u$. A base data set $\mathcal{D}_\text{base}$ is formed with training images and the corresponding labels of the base classes from all the base institutions.
\begin{align}
    \mathcal{D}_\text{base}=
\bigcup_{u\in \mathcal{U}_\text{base}}
\bigcup_{I\in \mathcal{I}^\text{train}_u} 
\bigcup_{c\in \mathcal{C}_\text{base}}\{~(I,~M(I, c))~\}.
\end{align}
Similarly, test images and the corresponding labels of novel classes from all base institutions form another data set,
\begin{align}
    \mathcal{D}_\text{base\_ins\_novel\_cls}=
\bigcup_{u\in \mathcal{U}_\text{base}}
\bigcup_{I\in \mathcal{I}^\text{test}_u} 
\bigcup_{c\in \mathcal{C}_\text{novel}}\{~(I,~M(I, c))~\},
\end{align}
together with the data set of images from the novel institutions and the novel classes' labels,
\begin{align}
    \mathcal{D}_\text{novel\_ins\_novel\_cls}=
\bigcup_{u\in \mathcal{U}_\text{novel}}
\bigcup_{I\in \mathcal{I}_u}\bigcup_{c\in \mathcal{C}_\text{novel}}\{~(I,~M(I, c))~\},
\end{align}
a novel data set is built focusing on novel classes:
\begin{align}
    \mathcal{D}_\text{novel}=\mathcal{D}_\text{base\_ins\_novel\_cls}~\cup~\mathcal{D}_\text{novel\_ins\_novel\_cls}.
\end{align}

Therefore, the cross-institution segmentation task aims to train a model on the base data set $\mathcal{D}_\text{base}$ and generalise to the novel data set $\mathcal{D}_\text{novel}$ which contains novel classes on both base and novel institutions.
Specifically, following the few-shot setting described in \cite{roy2020squeeze} and \cite{ ouyang2020self}, the model is tasked to segment a novel class $c\in\mathcal{C}_{novel}$ in a query image $I^q_u$ acquired from a novel institution $u\in\mathcal{U}_{novel}$ with only $K$ support examples $\{(I_{{u^s},k}^s,~M(I_{{u^s},k}^s, c))\}_{k=1}^{K}$ from the same or different institutions. The predicted query mask $\hat{M}(I^q_u, c)$ is compared with the label $M(I^q_u, c)$ with a segmentation metric such as the Dice score~\citep{sudre2017generalised}. Such evaluation procedure is named an \emph{episode} with $K$-shot, which is also detailed in Algorithm~\ref{alg:test}.

\begin{algorithm}
\caption{\label{alg:test}Evaluation Procedure}
\SetKwInOut{Input}{Input}\SetKwInOut{Output}{Output}
\Input{Neural network $\phi_{\theta}$ with parameters $\theta$\\
Images $\mathcal{I}_u^\text{test}$ for all base institutions $u\in \mathcal{U}_{base}$\\
Images $\mathcal{I}_u$ for all novel institutions $u\in \mathcal{U}_{novel}$\\
Masks for all novel classes in $\mathcal{C}_\text{novel}$
}
\Output{Dice scores for all, novel, and base institutions: \text{ALL\_DICE}, \text{NOVEL\_DICE}, and \text{BASE\_DICE}
}

$\text{ALL\_DICE\_SUM} = 0$\\
$\text{ALL\_DICE\_COUNT} = 0$\\
$\text{NOVEL\_DICE\_SUM} = 0$\\
$\text{NOVEL\_DICE\_COUNT} = 0$\\
$\text{BASE\_DICE\_SUM} = 0$\\
$\text{BASE\_DICE\_COUNT} = 0$\\
\For{$I^q \in \bigcup_{u\in\mathcal{U}_{novel}}\mathcal{I}_u$}{
    \For{$u \in \mathcal{U}_{novel}$}{
        Sample $I^s \in \mathcal{I}_u$, such that $I^s\neq I^q$\\
        \For {$c \in \mathcal{C}_\text{novel}$}{
            Denote the mask of $c$ in $I^s$ as $M^s(c)$\\
            Denote the mask of $c$ in $I^q$ as $M^q(c)$\\
            Predict $\hat{M}^q(c) = \phi_{\theta}(I^q, I^s, M^s(c))$\\
            Evaluate $dice=\text{Dice}(\hat{M}^q(c), M^q(c))$\\
            Update $\text{ALL\_DICE\_SUM} \mathrel{{+}{=}} dice$\\
            Update $\text{ALL\_DICE\_COUNT} \mathrel{{+}{=}} 1$\\
            Update $\text{NOVEL\_DICE\_SUM} \mathrel{{+}{=}} dice$\\
            Update $\text{NOVEL\_DICE\_COUNT} \mathrel{{+}{=}} 1$\\
        }
    }
    \For{$u \in \mathcal{U}_{base}$}{
        Sample $I^s \in \mathcal{I}_u^{test}$\\
        \For {$c \in \mathcal{C}_\text{novel}$}{
        Denote the mask of $c$ in $I^s$ as $M^s(c)$\\
            Denote the mask of $c$ in $I^q$ as $M^q(c)$\\
            Predict $\hat{M}^q(c) = \phi_{\theta}(I^q, I^s, M^s(c))$\\
            Evaluate $dice=\text{Dice}(\hat{M}^q(c), M^q(c))$\\
            Update $\text{ALL\_DICE\_SUM} \mathrel{{+}{=}} dice$\\
            Update $\text{ALL\_DICE\_COUNT} \mathrel{{+}{=}} 1$\\
            Update $\text{BASE\_DICE\_SUM} \mathrel{{+}{=}} dice$\\
            Update $\text{BASE\_DICE\_COUNT} \mathrel{{+}{=}} 1$\\
        }
    }
}
Compute $\text{ALL\_DICE} = \frac{\text{ALL\_DICE\_SUM}}{\text{ALL\_DICE\_COUNT}}$\\
Compute $\text{NOVEL\_DICE} = \frac{\text{NOVEL\_DICE\_SUM}}{\text{NOVEL\_DICE\_COUNT}}$\\
Compute $\text{BASE\_DICE} = \frac{\text{BASE\_DICE\_SUM}}{\text{BASE\_DICE\_COUNT}}$\\
\end{algorithm}

\section{Method}
\label{sec:method}

\subsection{Episodic Few-shot Training}
\label{sec:episodic}
This work adopts the common episodic training paradigm (detailed in Algorithm~\ref{alg:train}) that simulates the few-shot task during training. Each \emph{episode} consists of query $(I^q, M(I^q,c))$ and $K$ support $\{(I_k^s, M(I_k^s,c))\}_{k=1}^K$ image-label pairs, for a base class $c$ sampled from $C_\text{base}$.
In this work, $K=1$ during training and the model is trained to predict the query mask $M^q(c)$ given the query image $I^q$ and one support image-label pair, denoted as $(I^s, M^s(c))$.

\begin{algorithm}
\caption{\label{alg:train}Episodic Training Procedure}
\SetKwInOut{Input}{input}\SetKwInOut{Output}{output}
\Input{Neural network $\phi_{\theta}$ with parameters $\theta$.\\
Learning rate $\alpha$.\\
Images $\mathcal{I}_u^\text{train}$ for all base institutions $u\in \mathcal{U}_{base}$. \\
Masks for all base classes in $\mathcal{C}_\text{base}$. 
}
\Output{Trained model $\phi_{\theta}$.}
\While{Training}{
    Sample $c \sim \mathcal{C}_{base}$\\
    Sample $I^q,I^s \in \bigcup_{u\in\mathcal{U}_{base}}\mathcal{I}_u^\text{train}$, such that $I_s\neq I_q$\\
    Denote the mask of $c$ in $I^s$ as $M^s(c)$\\
    Denote the mask of $c$ in $I^q$ as $M^q(c)$\\
    Predict $\hat{M}^q(c) = \phi_{\theta}(I^q, I^s, M^s(c))$\\
    Compute loss: $L$\\
    Update model parameters: $\theta = \theta - \alpha\nabla_{\theta}L$
}
\end{algorithm}

\subsection{Prototypical Network}
A prototypical network~\citep{dong2018few} first defines a prototype feature vector per class, extracted from the embedded features of the labelled voxels, in the support image, corresponding to the class. The feature vector is then compared with the query image voxel-wise in the embedded feature space for segmentation prediction.

Specifically, the query and support images, denoted by $I^s$ and $I^q$ respectively, are encoded by a shared feature extractor into support and query feature maps of the same shape, $F^s$ and $F^q$, respectively.
The class prototype $h_c$ and the background prototype $h_0$ are then derived by averaging $F^s$ over voxels of the class $c$ (where label $M^s(c)$ equals $1$) and background (where label $M^s(c)$ equals $0$), respectively, i.e., $h_c = f(F^s, M^s(c))$ and $h_0 = f(F^s, 1-M^s(c))$, where
\begin{align}\label{eq:prototype}
    f(F, M) = \frac{\sum\limits_{(x,y,z)} F_{(x,y,z)} M_{(x,y,z)}}{\sum\limits_{(x,y,z)}{M_{(x,y,z)}}},
\end{align}
with $(x,y,z)$ iterating over all voxels in $F$ along x, y and z axes.

The similarity between each query voxel and class (or background) is calculated through cosine similarity between the voxel feature map $F^q$ and the class prototype feature vector $h_c$ (or $h_0$ for background):
\begin{align}
    \text{sim}(\star)&=\frac{F^q_{(x,y,z)}\cdot h_\star}{\|F^q_{(x,y,z)}\|_2~\|h_\star\|_2},
\end{align}
where $\star\in\{c,0\}$ represents the class or background, and $\cdot$ represents the dot product between vectors.

\subsection{Local Prototypical Network}
\label{sec:local_prototypical_network}

To extract location-sensitive local prototypes \citep{yu2021location}, images of spatial size $W\times H \times D$ are partitioned into overlapping windows $g \in G$ of size $\alpha_w W \times \alpha_h H \times \alpha_d D$ with the equidistant spacing between window centres being half of the window size.
As shown in Fig.~\ref{fig:3d_con_align}, for each window $g \in G$, two local prototype feature vectors $h_c^g$ and $h_0^g$ are calculated via Eq.~\eqref{eq:prototype} by iterating $(x,y,z)$ over the voxels inside the window $g$:
$h_c^g = f(F^s, M^s(c), g)$ and $h_0^g = f(F^s, 1-M^s(c), g)$, where
\begin{align}\label{eq:local_prototype}
    f(F, M, g) = \frac{\sum\limits_{(x,y,z)\in g} F_{(x,y,z)} M_{(x,y,z)}}{\sum\limits_{(x,y,z)\in g}{M_{(x,y,z)}}}.
\end{align}

For each query voxel $(x,y,z)$, $\mathcal{G}_{(x,y,z)}$ denotes the set of all windows that contains the voxel: $\mathcal{G}_{(x,y,z)}=\{g \mid (x,y,z) \in g\}$.
The local similarity between this voxel and the class (or background) is then calculated using the maximum cosine similarity over windows $g\in\mathcal{G}_{(x,y,z)}$ with the corresponding local prototype feature vectors $\{h_c^g\}_{g\in\mathcal{G}_{(x,y,z)}}$ and $\{h_0^g\}_{g\in\mathcal{G}_{(x,y,z)}}$:
\begin{align}
    \text{sim}(\star)=\max\limits_{g\in\mathcal{G}_{(x,y,z)}}\frac{F^q_{(x,y,z)}\cdot h_\star^g}{\|F^q_{(x,y,z)}\|_2~\|h_\star^g\|_2}
\end{align}
with $\star\in\{c,0\}$ representing the class or background.
The foreground/background probability map is derived by:
\begin{align}
    \hat{M}^q(\star)=\frac{\exp(\text{sim}(\star))}{\exp(\text{sim}(0))+\exp(\text{sim}(c))}  
\end{align}
with $\star\in\{c,0\}$ representing the class and background.
The model is trained to minimise the Dice loss between the predicted and ground-truth mask:
\begin{align}
    L_{\text{fewshot}}=1 - \sum_{\star\in\{0, c\}}\frac{2\sum_{(x,y,z)} \hat{M}^q(\star)_{(x,y,z)} M^q(\star)_{(x,y,z)}}{\sum_{(x,y,z)} \hat{M}^q(\star)_{(x,y,z)}^2 + \sum_{(x,y,z)} M^q(\star)_{(x,y,z)}^2}
    \label{eq:few_shot_loss}
\end{align}
where $M^q(0)=1-M^q(c)$

\subsection{Spatial Registration Mechanism}
\label{sec:align}

\begin{figure*}[h!]
\centering
\includegraphics[scale=.45]{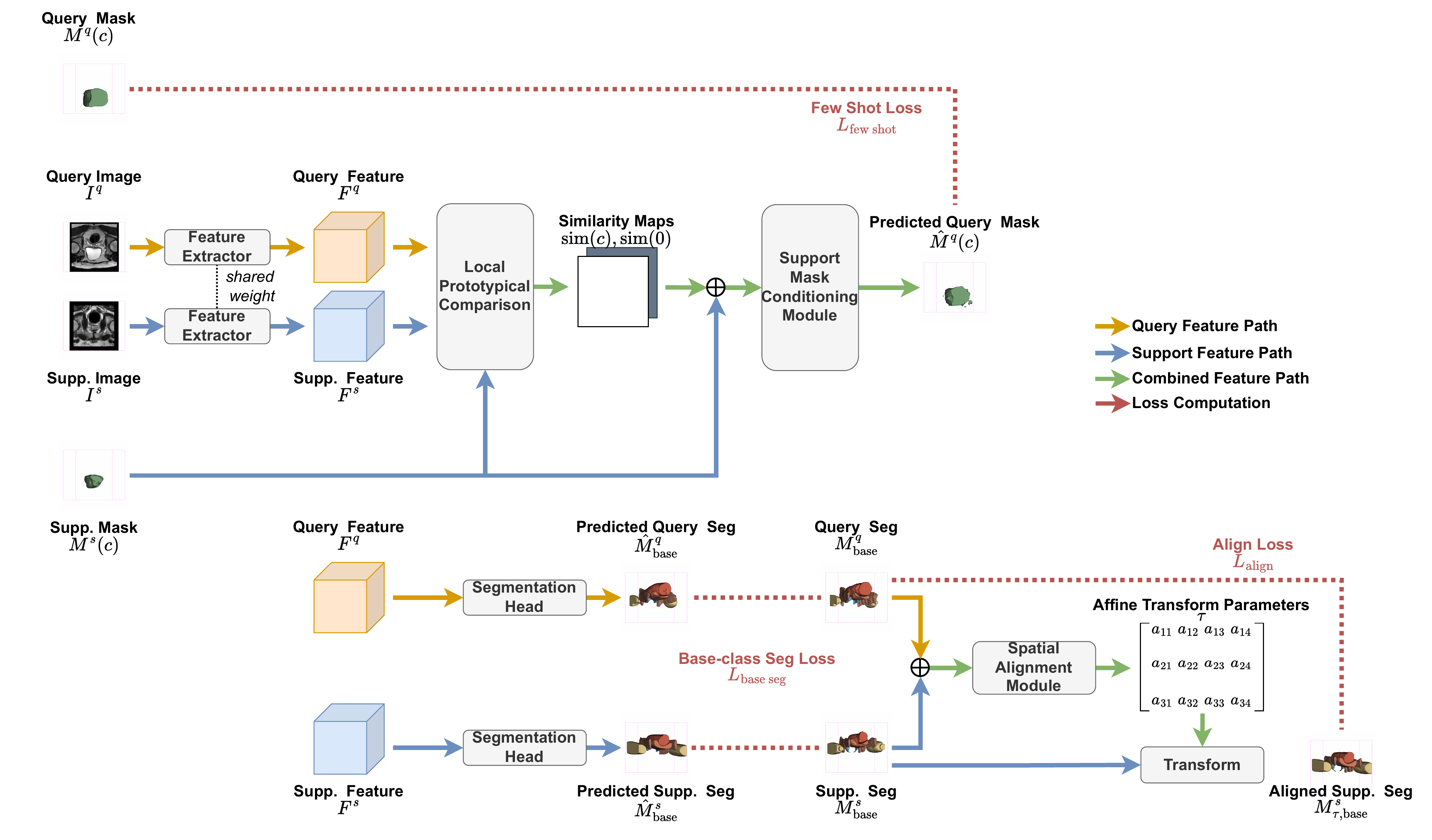}
\caption{The training procedure of the proposed method. The model is trained to minimise the sum of few-shot loss (Eq.~\ref{eq:few_shot_loss}), base class segmentation loss (Eq.~\ref{eq:base_seg_loss}) and alignment loss (Eq.~\ref{eq:align_loss}).}
\label{fig:train}
\end{figure*}

As discussed in Section~\ref{sec:intro}, the differences in intra- and inter-institution variations pose challenges in the local prototypical network due to the varying image sizes, orientations, and voxel dimensions of the acquiring institution. The target structure in the query and support images can be distant (as in Fig.~\ref{fig:align_vis}). Therefore, they may not be included inside of the same or even adjacent windows, and this results in erroneous comparison between the query voxels of the structure, thus irrelevant prototype vectors.
However, while the absolute locations of target structures varies among images, the relative position between different structures remains consistent. This observation motivated spatial alignment of the query and the support images, $I^q$ and $I^s$, illustrated in Fig.~\ref{fig:3d_con_align}, before extracting the local prototype feature vectors. This spatial registration process is conjectured to alleviate the discrepancy between different institutions and therefore reduce the amount of cross-institution training data required for generalisation. 

In this work, we consider an affine transformation to account for the above-discussed spatial difference with potentially uncertainties due to variable imaging positioning, signal sampling and scanner calibration, although higher-degree transformation will also be of interest. Furthermore, to avoid repeated feature map extraction, we propose to apply the transformation directly on feature maps ($F^q$ and $F^s$), rather than on images ($I^q$ and $I^s$).

The affine transformation prediction consists of two stages.
Firstly, a shared base class segmentation head segments all base classes from the query and support feature maps, $F^q$ and $F^s$, respectively. The multi-class predictions are denoted as $\hat{M}^q_\text{base}$ and $\hat{M}^s_\text{base}$, for the query and support images, respectively. 
During evaluation, these predictions are concatenated and passed into the \emph{spatial alignment module}, illustrated in Fig. \ref{fig:3d_con_align}, which predicts an affine transformation matrix $\tau\in\mathbb{R}^{3\times4}$ of $12$ degrees of freedom. During training, as illustrated in Fig.~\ref{fig:train}, base classes segmentation masks $M^q_\text{base}$ and $M^s_\text{base}$ are used for alignment prediction.
Secondly, alignment $\tau$ is applied to the support feature map $F^s$ and the label $M_\text{base}^s$, to obtain the aligned support feature map $F_\tau^s=\tau\circ F^s$ and the aligned support label for all base classes $M_{\tau,\text{base}}^s=\tau\circ M_\text{base}^s$, respectively. 
These aligned feature maps are then used to generate the local prototypes (as detailed in Section~\ref{sec:local_prototypical_network}) for segmentation.

Two losses are defined for training the spatial registration mechanism. First, a Dice loss is defined for the multi-class segmentation:
\begin{align}
    L_\text{base seg}= L_\text{multi-class}(\hat{M}^q_\text{base}, M^q_\text{base}) + L_\text{multi-class}(\hat{M}^s_\text{base}, M^s_\text{base}),
    \label{eq:base_seg_loss}
\end{align}
where the multi-class Dice loss is defined as:
\begin{align}
    &L_\text{multi-class}(M,\hat{M})\nonumber\\
    =& 1-\sum_{\star\in\mathcal{C}_\text{base}\cup\{0\}}\frac{2\sum_{(x,y,z)} \hat{M}(\star)_{(x,y,z)} M(\star)_{(x,y,z)}}{\sum_{(x,y,z)} \hat{M}(\star)_{(x,y,z)}^2 + \sum_{(x,y,z)} M(\star)_{(x,y,z)}^2}
\end{align}
with $M(\star)_{(x,y,z)}$ and $\hat{M}(\star)_{(x,y,z)}$ representing the ground truth and the predicted probability of a base class or background $\star$ at $(x,y,z)$.

The second loss aims to optimise alignment by minimising the Dice loss between the query label $M^q_\text{base}$ and the aligned support label $M_{\tau,\text{base}}^s$ of all base classes:
\begin{align}
    L_\text{align}= L_\text{multi-class}(M^q_\text{base},M_{\tau,\text{base}}^s).
    \label{eq:align_loss}
\end{align}

In theory, the transformation could be applied the other way around, i.e. by applying the reverse alignment $\tau^{-1}$ to the query feature map $F^q$ resulting in the aligned query feature map $F_{\tau^{-1}}^q=\tau^{-1}\circ F^q$, to achieve spatial alignment between query and support features. A cycle-consistent two-way registration may also apply. However, once the aligned query mask $\hat{M}_{\tau^{-1}}^q$ is predicted, it needs to be inverted, in order to obtain the mask of the original query image $\hat{M}^q = \tau\circ \hat{M}_{\tau^{-1}}^q$. The mechanism was adopted in this work for its computational efficiency without additional resampling in practice.

\subsection{Support Mask Conditioning Module}
\label{sec:shape_aware}

During the prototype feature vector extraction using the novel class mask in the support image, the voxel-wise information may be made invariant to the transformation during the aggregation in Eq. \eqref{eq:prototype} and \eqref{eq:local_prototype}, which is designed for ``normalising'' cross-institution data, but may also result in large spatial variability in the prediction. Therefore, a simple yet effective \emph{support mask conditioning module} is proposed.

The support mask conditioning module takes as input a concatenation of the class similarity $\text{sim}(c)$, the background similarity $\text{sim}(0)$ and the aligned mask of the class in the support image $M_{\tau}^s(c)$, for the final prediction $\hat{M}^q(c)$. Unlike the multiplication of support mask to support features for prototype calculation (in Eq.~\eqref{eq:prototype} and \eqref{eq:local_prototype}), the direct use of support mask here provides a more direct route for the novel class information to the final segmentation task, similar to commonly designed shortcut layers for skipping networks.

\textcolor{black}{Different from 3D medical segmentation algorithms using mask predicted from downsampled image to provide context information for high-resolution patches, the proposed method uses the segmentation of query image with location and shape information of the support mask.}

\begin{figure*}[t]
\centering
\includegraphics[scale=.2]{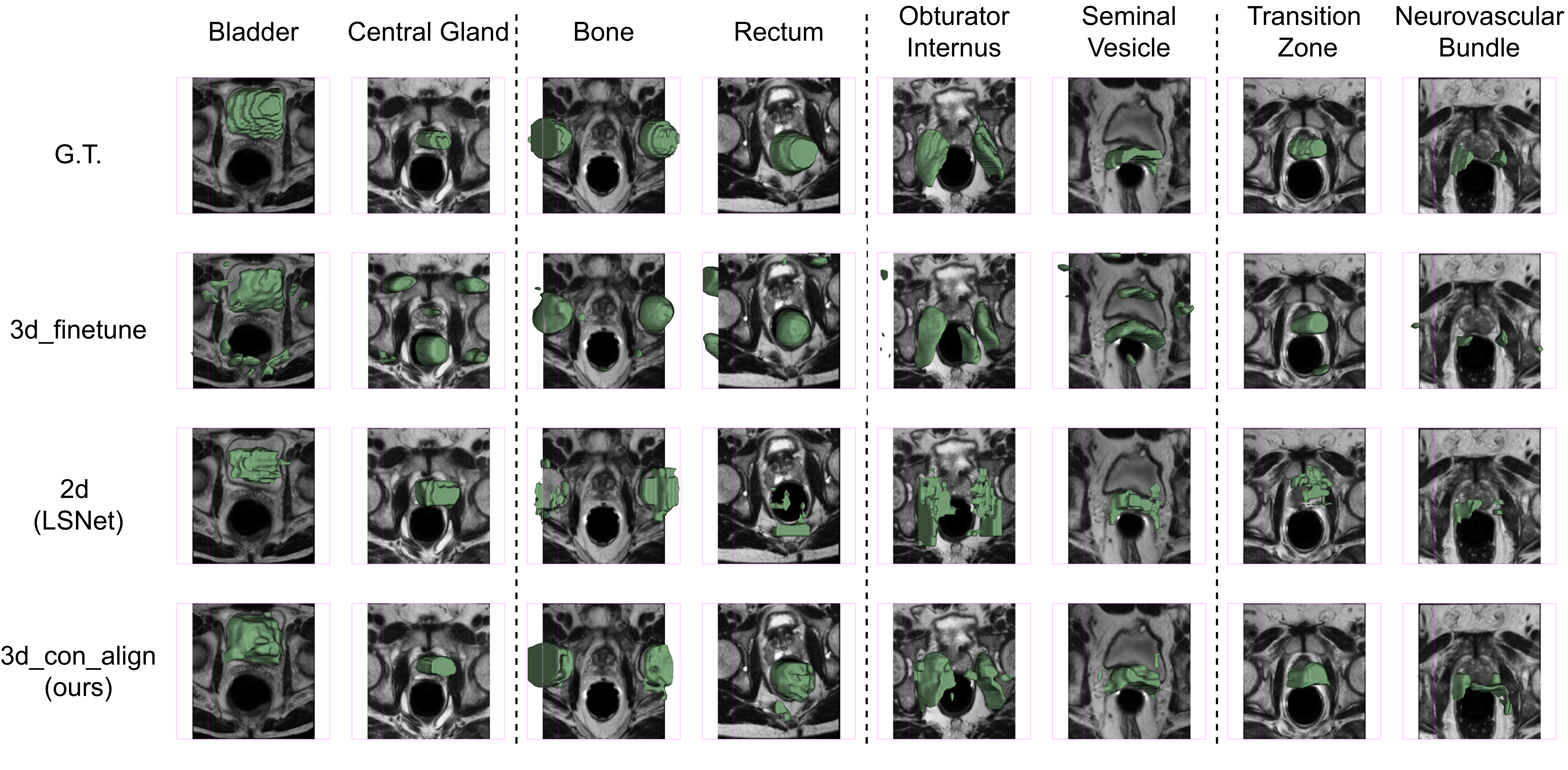}
\caption{Qualitative result achieved by `3d\_finetune' and `2d' baselines (detailed in Section~\ref{sec:baseline}) and our proposed method.}
\label{fig:vis}
\end{figure*}

\subsection{Loss}
\label{sec:loss}
Both the spatial registration mechanism and support mask conditioning module are trained with the original local prototypical network, with an overall training loss function used in this study as follows:
\begin{align}
    L=L_{\text{fewshot}} + L_{\text{base seg}} + L_{\text{align}}
\end{align}

\subsection{Multiple-shot Evaluation}

Due to memory limitation, the training was carried out in one-shot paradigm, i.e. $K=1$. During the evaluation, for the query image $I^q$ and each of the $K$ support images $I^s_k$, the base class segmentation is predicted from the base class segmentation head, denoted by $\hat{M}^q_\text{base}$ and $\hat{M}^s_{\text{base},k}$, respectively.
Among the $K$ support images, only the support image that is the most similar to the query image in terms of the cosine similarity on base class segmentation prediction, denoted by $I^s_{\hat{k}}$, is chosen to calculate the local prototypes in \eqref{eq:local_prototype}. Precisely, 
\begin{align}
\hat{k}=\arg\max\limits_{k}\frac{
\hat{M}^s_{\text{base}, k}\cdot\hat{M}^q_\text{base}
}{\|\hat{M}^s_{\text{base}, k}\|_2~\|\hat{M}^q_\text{base}\|_2},
\end{align}
where the dot product and norm are calculated on the flattened predictions.
The local prototypes are then calculated using the selected support example,
$h_c^g = f(F^s_{\hat{k}}, M(I^s_{\hat{k}}, c), g)$ and $h_0^g = f(F^s_{\hat{k}}, 1-M(I^s_{\hat{k}}, c), g)$ for each window $g$.

\section{Experiments}

\subsection{Data Set}
\begin{table}[]
    \centering
    \caption{\label{tab:institution}The number of images acquired from each institution.}
    \begin{tabular}{|c|ccccccc|c|}
    \hline
    &\multicolumn{7}{c|}{institution}&\\
    &1&2&3&4&5&6&7&total\\\hline
    \# of images&321&45&74&82&24&24&19&589\\\hline
    \end{tabular}
\end{table}

The data set includes 589 T2-weighted images acquired from the same number of patients collected by seven studies, INDEX~\citep{dickinson2013multi}, the SmartTarget Biopsy Trial~\citep{hamid2019smarttarget}, PICTURE~\citep{simmons2014picture}, TCIA Prostate3T~\citep{Prostate-3T}, Promise12~\citep{litjens2014evaluation}, TCIA ProstateDx (Diagnosis)~\citep{ProstateDx} and the Prostate MR Image Database~\citep{ProstateMRI}. Further details are reported in the respective study references. 

These images were divided into seven subsets based on the acquiring institution. The number of images acquired from each institution is anonymously summarised in Table~\ref{tab:institution}. The cross-institution imaging protocols contain multiple scanners (two manufacturers with a mixed 1.5 and 3T field strengths), varying field-of-view and anisotropic voxels, in-plane voxel dimensions ranging between 0.3 and 1.0 mm and out-of-plane spacing between 1.8 and 5.4 mm.

For each image, eight anatomical structures of planning interest, including bladder, bone, central gland, neurovascular bundle, obturator internus, rectum, seminal vesicle, transition zone were labelled (as shown in Fig.\ref{fig:vis}). All segmentations were manually annotated by eight biomedical imaging researchers, with experience ranging from 2 to 10 years in the annotation of medical image data, each annotating a mixed-institution subset using an institution-stratified sampling. Each annotation has been reviewed at least once.

The full segmentation masks and the derived intensity arrays from T2-weighted sequences, after pre-processing, used to produce the results in this study, are available at \dataurl.

The eight lower-pelvic structures were randomly divided into four folds as shown in Table~\ref{tab:class_fold}. In a cross-validation experiment, the classes contained in each fold were considered as novel classes, the other three folds representing base classes.
The images of each institution $u$ were then randomly sampled into training and testing subsets in a 3:1 ratio. A further validation set was formed with 12 images, 2 from each institution, randomly chosen from the novel data set $\mathcal{D}_{novel}$. Those images were excluded from the testing. Unless otherwise specified, the same data partitioning was used for all the results presented. All statistical conclusions are reported using paired Student's t-tests at the significance level of $\alpha_{t-test}$ = 0.05.

\begin{table}[]
    \centering
    \caption{\label{tab:class_fold}The eight structures are randomly divided into 4 folds.}
    \begin{tabular}{|c|c|}
    \hline
           & structures \\\hline
    fold 1 & bladder, central gland\\\hline
    fold 2 & bone, rectum\\\hline
    fold 3 & obturator internus, seminal vesicle\\\hline
    fold 4 & transition zone, neurovascular bundle\\\hline
    \end{tabular}
\end{table}

\begin{figure}[h]
\centering
\includegraphics[scale=.45]{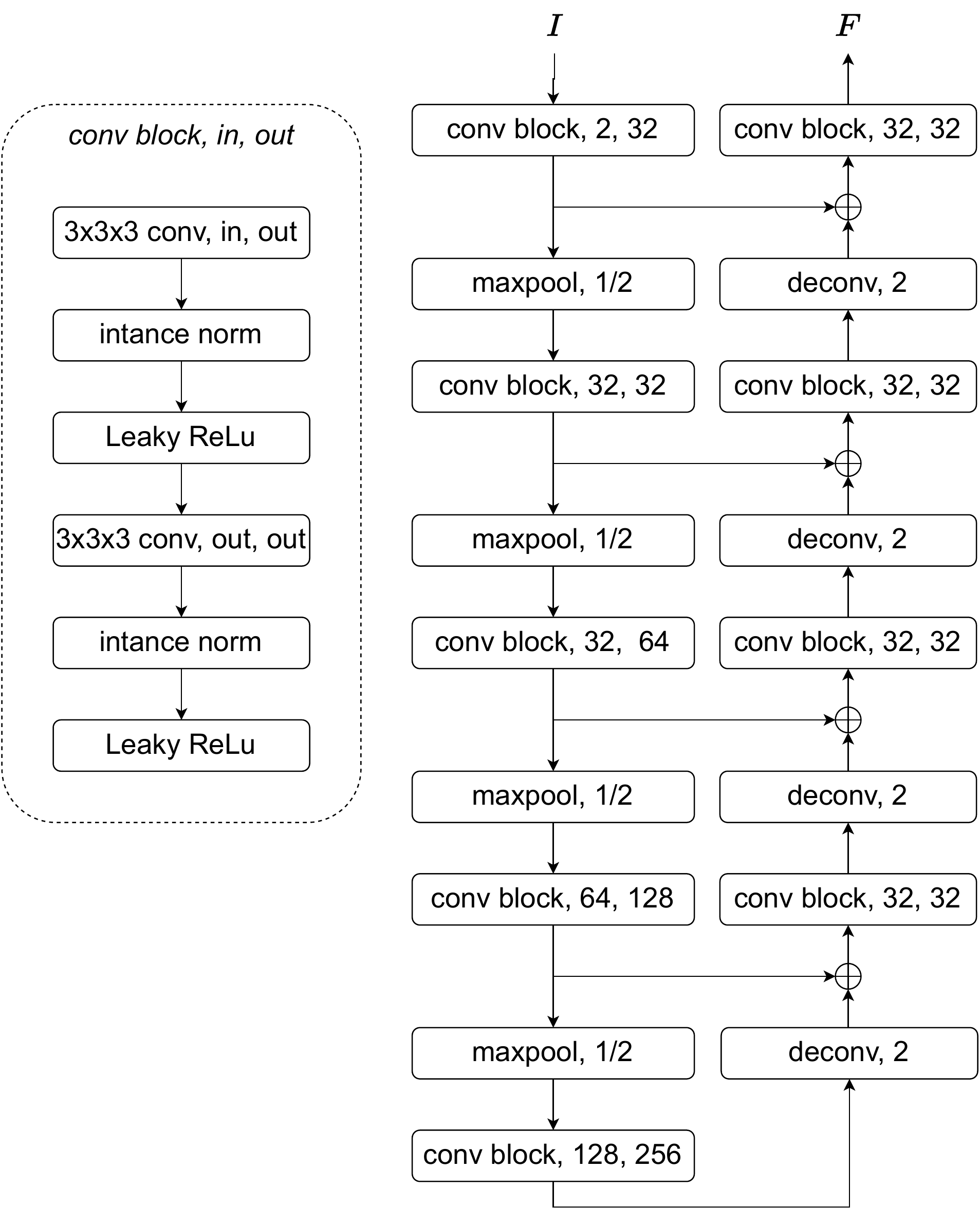}
\caption{Feature extractor architecture.}
\label{fig:unet_architecture}
\end{figure}

\begin{figure}[h]
\centering
\includegraphics[scale=.4]{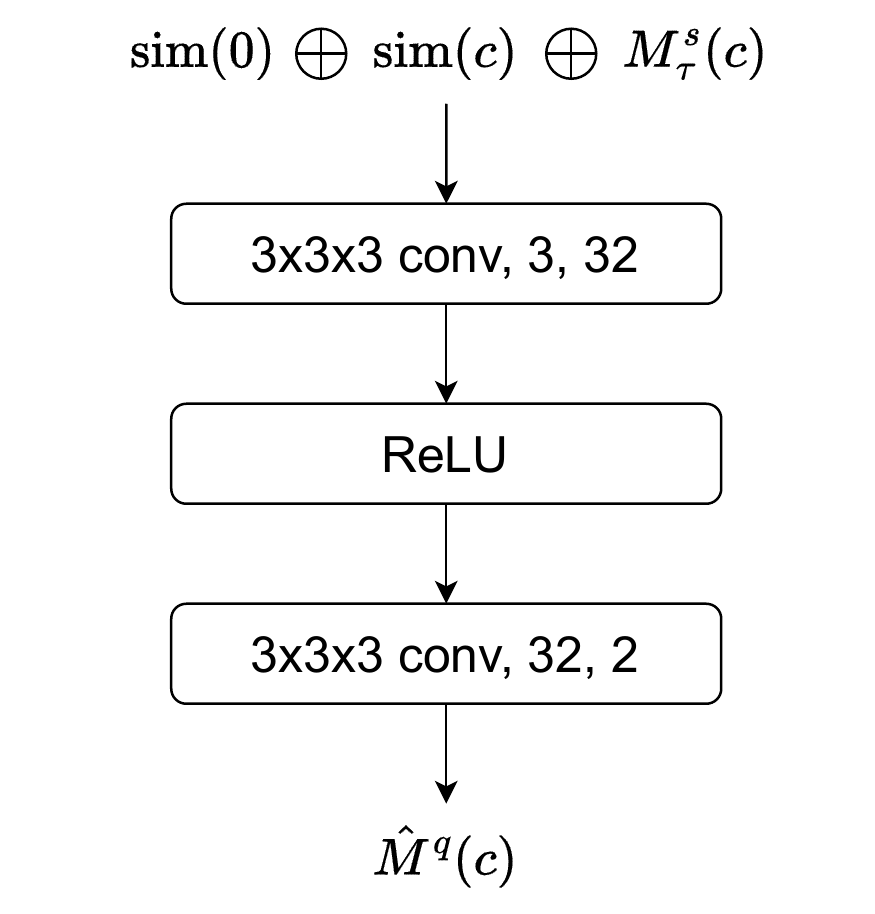}
\caption{Support mask conditioning module architecture.}
\label{fig:con_architecture}
\end{figure}

\begin{figure}[h]
\centering
\includegraphics[scale=.45]{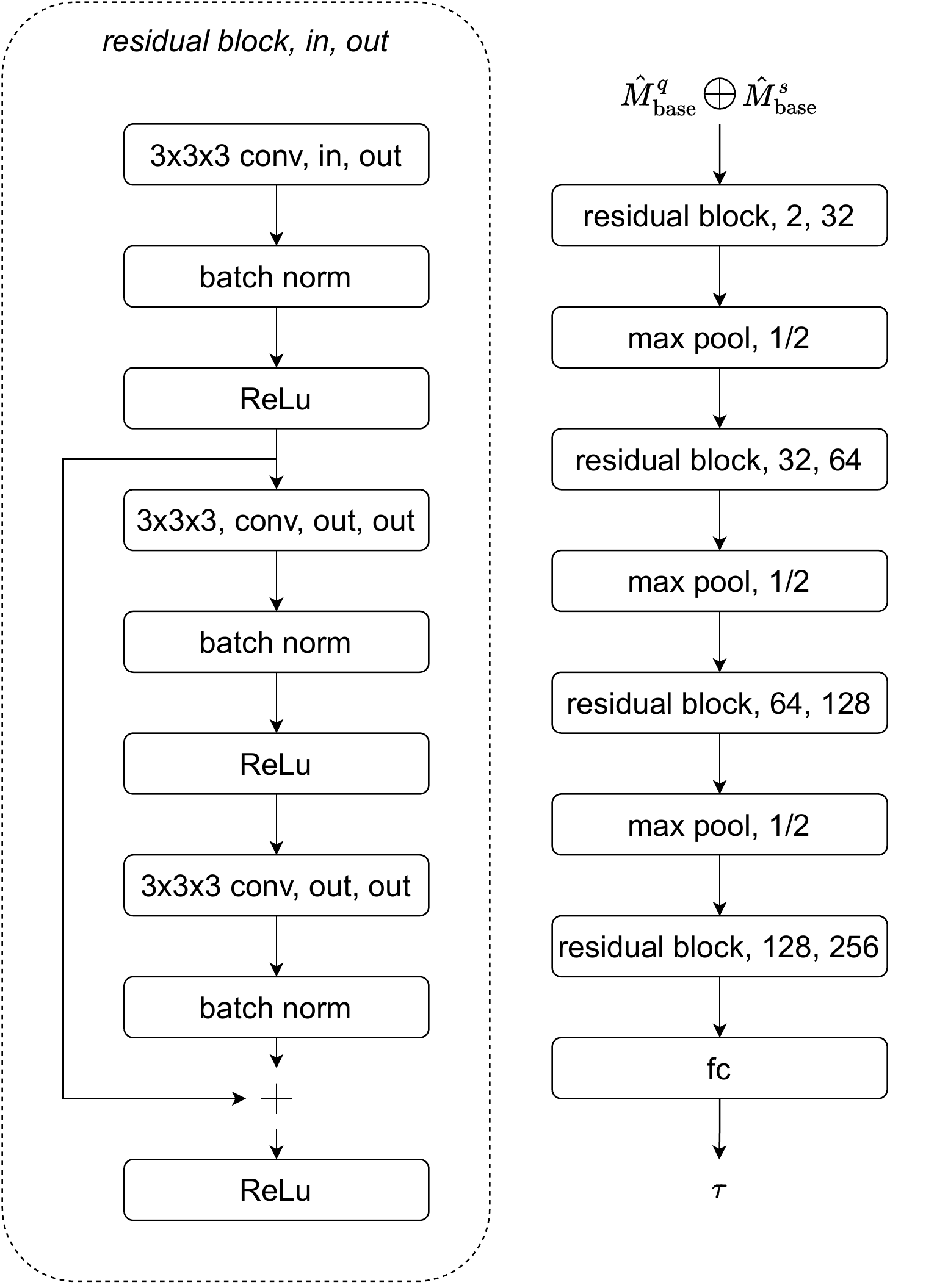}
\caption{Spatial alignment module architecture.}
\label{fig:align_architecture}
\end{figure}

\subsection{Implementation Details}
\label{sec:implementation_details}

\begin{figure*}[h!]
    \subfigure[]{
    \label{fig:3d}
    \includegraphics[scale=.4]{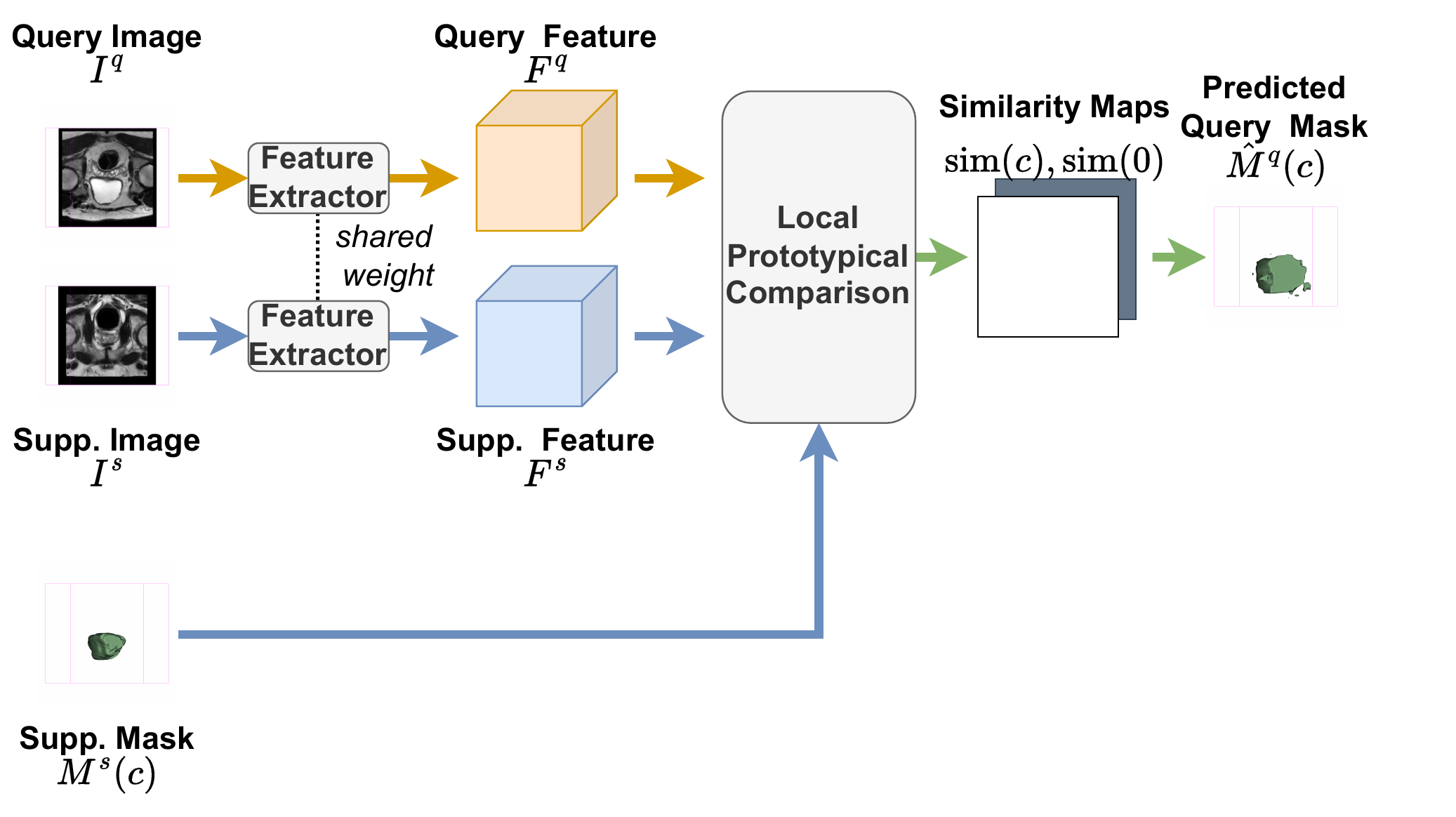}
    }
    \subfigure[]{
    \label{fig:3d_con}
    \includegraphics[scale=.4]{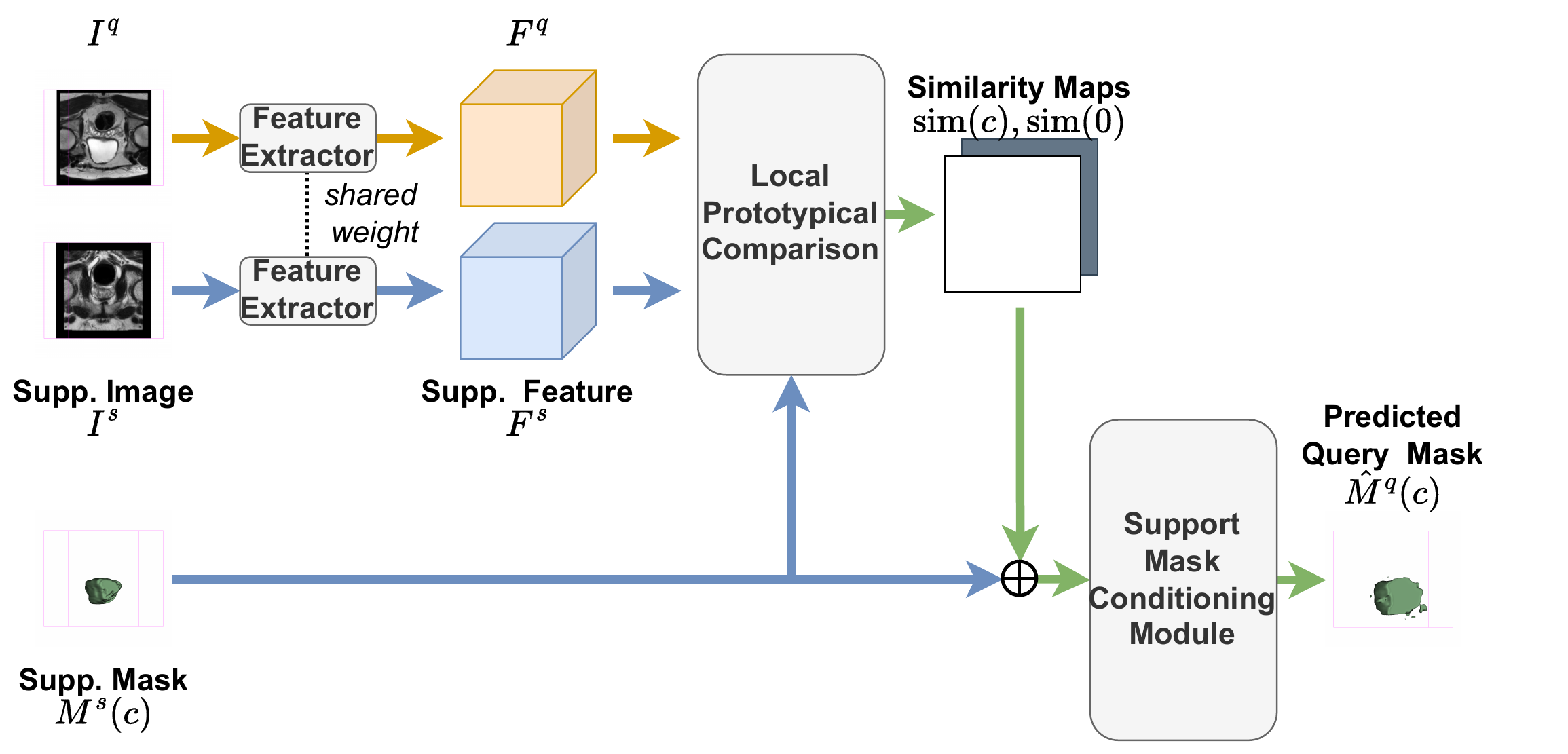}
    }
    \subfigure[]{
    \label{fig:3d_align}
    \includegraphics[scale=.4]{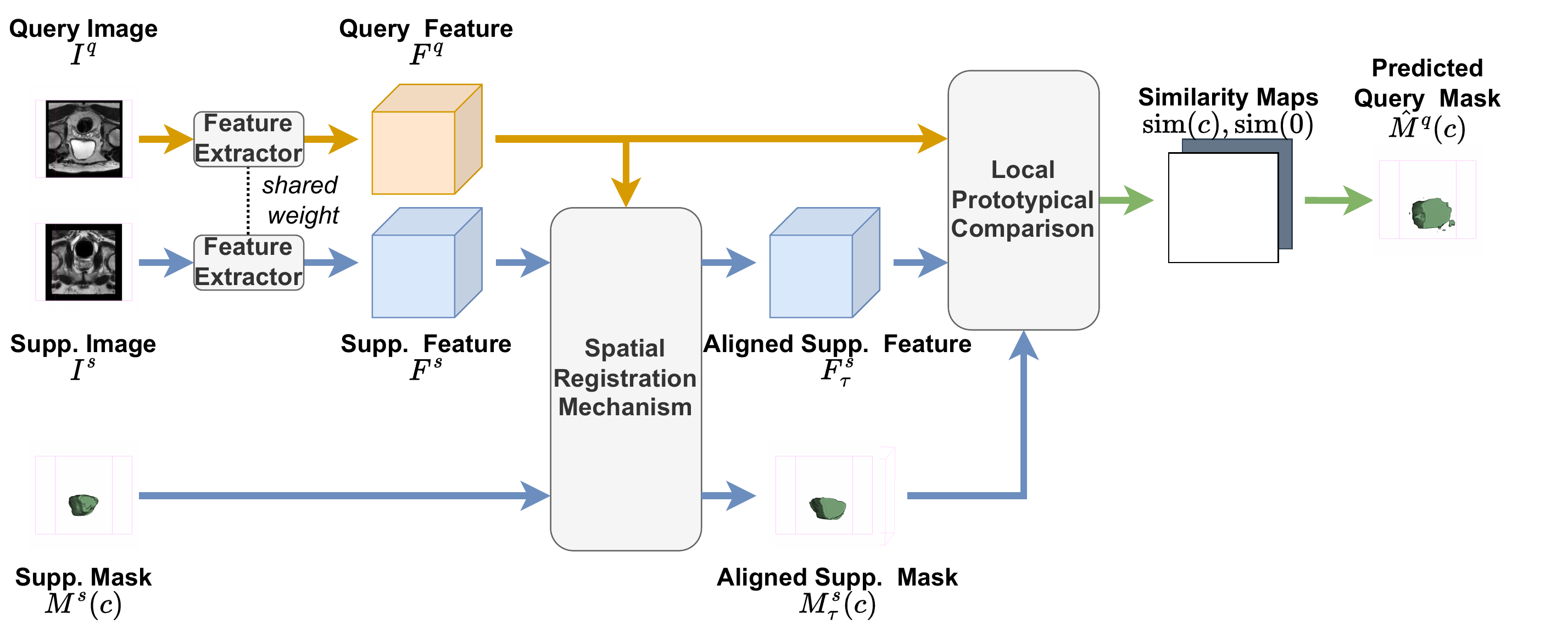}
    }
\caption{Illustration of the three model variations (a) `3d' (b) `3d\_con' (c) `3d\_align' experimented for ablation studies (Section~\ref{sec:ablation_studies})}
\end{figure*}

All images were normalised, resampled and centre-cropped to an image size of $256\times256\times48$, with a voxel dimension of $0.75\times0.75\times2.5$ during pre-processing. Random rotation, translation and scaling were adopted for data augmentation during training.

The best training episode was chosen based on the performance on the validation sets. 
During evaluation, all images from the novel institution were considered query images for evaluation for each of the fold-specified novel classes. 
A binary Dice score for this novel class was calculated for each query image based on a sampled support image from each of the seven institutions, excluding the query. As described in Algorithm~\ref{alg:test}, the results were reported when support images from 1) all institutions, 2) base institutions, and 3) novel institutions. The institution where the support image comes from is denoted as `support ins'.

A 3D UNet was adopted as the feature extractor, whose architecture is detailed in Fig.~\ref{fig:unet_architecture}. $\alpha_w=\alpha_h=\frac{1}{8}$ and $\alpha_d=1$ were selected for local prototypical comparison as detailed in Section~\ref{sec:local_prototypical_network}. As shown in Fig.~\ref{fig:con_architecture}, the support mask conditioning module was made up of two convolutional layers. For the spatial registration mechanism, the base class segmentation head was a single convolutional layer, and the spatial alignment module was a GlobalNet~\citep{hu2018label} as specified in Fig.~\ref{fig:align_architecture}. The models were trained using an Adam optimiser starting at a learning rate of $10^{-4}$ with a minibatch size of 1. The implementation code has been released at \codeurl.

\subsection{Compared Baseline Networks}
\label{sec:baseline}
For comparison, we report the results of the following baseline networks. 
\begin{enumerate}
    \item The `3d\_finetune' network - The `3d\_finetune' baseline implemented the same UNet as the feature extractor. It was pre-trained on the base data set to segment all base classes for $100$ epochs using an Adam optimiser starting at a learning rate of $10^{-4}$. During evaluation, for each query, the pre-trained model is fine-tuned on the support images for $10$ iterations before testing. This baseline provides a reference as a ``lower-bound'' performance, using a simple transfer learning strategy.
    \item The `2d' network - LSNet~\citep{yu2021location} was adopted as the 2D episodic baseline. It adopted the same local-prototype comparison approach as detailed in Section~\ref{sec:local_prototypical_network} but using a 2D backbone based on ResNet-50 pre-trained on ImageNet, instead of 3D networks. To the best of our knowledge, this is the prototypical network closest to our work that has been proposed for medical image segmentation.
    \item BiGRU~\citep{kim2021bidirectional} - Another recent few-shot medical segmentation method with a UNet-like network for 2D slice prediction and a bidirectional gated recurrent unit (GRU) for adjacent slices consistency.
    \item The `localnet (unsupervised)' network for multi-atlas segmentation - A non-rigid registration network, LocalNet~\citep{hu2018weakly}, trained on the base data set images with no organ label supervision. 
    Given a support-query pair, the model is trained to predict a dense displacement field that warps the support towards the query. For n-shot evaluation, $n$ dense displacement fields are predicted, each registering a support example towards the query. $n$ query mask predictions are derived by warping each support mask using the corresponding predicted dense displacement field. The final prediction is made through majority voting by the $n$ query mask predictions.
    The implementation was based on the open-source repository MONAI~\citep{cardoso2022monai}.
    \item The `localnet (supervised)' network for multi-atlas segmentation  - A non-rigid registration network, LocalNet~\citep{hu2018weakly}, trained on the base data set with masks from all classes including novel classes. Similar to the `localnet(unsupervised)' network, the query mask is the warped support mask using the registration-predicted support-to-query dense displacement field. The implementation was based on the open-source repository MONAI~\citep{cardoso2022monai}.
    \item The `3d\_supervised' network - A fully supervised 3D model was trained on the base data set images with masks from all classes. The results on the novel institution images are reported as an ``upper-bound'' performance.
\end{enumerate}

\newcommand{\Dicethree}{\small
\begin{tabular}{|c|c|ccccc|ccccc|c|}%
\hline%
\multirow{2}{*}{model}&support&\multicolumn{5}{c|}{Dice (\%)}&\multicolumn{5}{c|}{95\% Hausdorff distance (mm)}&\#param\\%
&ins&fold1&fold2&fold3&fold4&mean&fold1&fold2&fold3&fold4&mean&(million)\\%
\hline%
\multirow{4}{*}{3d\_finetune}&all&48.07&44.78&43.92&29.25&41.50&61.47&69.78&60.33&57.02&62.15&\multirow{4}{*}{5.75}\\%
&base&45.74&39.81&41.69&26.02&38.32&63.22&73.42&62.33&59.24&64.55&\\%
&novel&62.03&74.56&57.31&48.63&60.63&50.98&47.96&48.32&43.72&47.74&\\%
&$\Delta$&26.26\%&46.60\%&27.25\%&46.48\%&36.80\%&{-}24.02\%&{-}53.09\%&{-}28.99\%&{-}35.51\%&{-}35.21\%&\\%
\hline%
\multirow{4}{*}{BiGRU}&all&44.05&43.08&29.61&20.97&34.43&23.70&33.67&23.60&22.83&25.95&\multirow{4}{*}{38.85}\\%
&base&44.05&42.74&29.21&20.62&34.15&23.73&34.06&23.85&23.20&26.21&\\%
&novel&44.03&45.11&32.01&23.12&36.07&23.49&31.30&22.12&20.60&24.38&\\%
&$\Delta$&{-}0.06\%&5.26\%&8.75\%&10.80\%&5.30\%&{-}1.04\%&{-}8.82\%&{-}7.83\%&{-}12.59\%&{-}7.52\%&\\
\hline%
\multirow{4}{*}{2d}&all&49.23&44.51&36.34&27.44&39.38&24.03&30.86&29.51&30.77&28.79&\multirow{4}{*}{23.63}\\%
&base&48.80&42.92&35.52&26.83&38.52&24.12&31.93&30.30&31.54&29.47&\\%
&novel&51.80&54.07&41.28&31.08&44.56&23.52&24.39&24.83&26.13&24.72&\\%
&$\Delta$&5.80\%&20.62\%&13.94\%&13.68\%&13.55\%&{-}2.54\%&{-}30.92\%&{-}22.03\%&{-}20.72\%&{-}19.24\%&\\%
\hline%
\multirow{4}{*}{3d}&all&47.79&52.06&34.81&26.16&40.20&31.60&30.68&32.95&31.71&31.73&\multirow{4}{*}{5.75}\\%
&base&47.89&51.07&34.12&25.44&39.63&31.69&31.03&33.35&32.19&32.07&\\%
&novel&47.22&58.00&38.96&30.49&43.67&31.07&28.54&30.54&28.81&29.74&\\%
&$\Delta$&{-}1.42\%&11.94\%&12.44\%&16.58\%&9.25\%&{-}2.00\%&{-}8.73\%&{-}9.20\%&{-}11.73\%&{-}7.82\%&\\%
\hline%
\multirow{4}{*}{3d\_con}&all&58.34&46.11&40.43&30.94&43.96&27.31&27.15&27.33&26.38&27.04&\multirow{4}{*}{5.75}\\%
&base&58.25&43.58&39.87&29.80&42.88&27.69&28.34&28.03&27.06&27.78&\\%
&novel&58.84&61.30&43.81&37.78&50.43&25.04&20.02&23.14&22.30&22.63&\\%
&$\Delta$&1.00\%&28.91\%&8.99\%&21.12\%&14.98\%&{-}10.56\%&{-}41.54\%&{-}21.12\%&{-}21.31\%&{-}22.76\%&\\%
\hline%
\multirow{4}{*}{3d\_align}&all&50.80&51.56&32.39&30.89&41.41&27.33&36.29&32.63&29.93&31.55&\multirow{4}{*}{27.34}\\%
&base&50.60&51.17&31.25&30.82&40.96&27.35&36.82&33.01&30.42&31.90&\\%
&novel&51.97&53.88&39.22&31.33&44.10&27.20&33.15&30.34&26.96&29.41&\\%
&$\Delta$&2.64\%&5.04\%&20.31\%&1.62\%&7.12\%&{-}0.55\%&{-}11.08\%&{-}8.78\%&{-}12.84\%&{-}8.46\%&\\%
\hline%
\multirow{4}{*}{3d\_con\_align}&all&59.36&60.38&45.73&37.60&50.77&22.66&29.63&27.92&26.41&26.66&\multirow{4}{*}{27.35}\\%
&base&59.11&59.80&45.06&37.28&50.31&22.69&30.30&28.38&26.89&27.06&\\%
&novel&60.86&63.87&49.77&39.54&53.51&22.49&25.63&25.19&23.54&24.21&\\%
&$\Delta$&2.89\%&6.37\%&9.47\%&5.70\%&5.98\%&{-}0.91\%&{-}18.21\%&{-}12.65\%&{-}14.26\%&{-}11.79\%&\\%
\hline%
\multirow{2}{*}{localnet}&all&25.64&38.99&21.83&11.20&24.42&28.74&27.21&24.78&26.97&26.93&\multirow{4}{*}{5.94}\\%
&base&24.16&36.43&21.03&10.35&22.99&29.43&28.48&25.39&27.89&27.80&\\%
\multirow{2}{*}{(unsupervised)}&novel&34.52&54.37&26.61&16.34&32.96&24.60&19.55&21.10&21.47&21.68&\\%
&$\Delta$&30.01\%&33.00\%&20.98\%&36.67\%&30.25\%&{-}19.63\%&{-}45.70\%&{-}20.32\%&{-}29.93\%&{-}28.22\%&\\%
\hline%
\multirow{2}{*}{localnet}&all&72.70&72.14&52.84&42.65&60.08&19.36&22.81&21.78&24.00&21.99&\multirow{4}{*}{5.94}\\%
&base&71.80&70.11&51.21&41.65&58.69&20.29&24.69&23.07&25.11&23.29&\\%
\multirow{2}{*}{(supervised)}&novel&78.11&84.33&62.61&48.61&68.42&13.77&11.57&14.03&17.29&14.17&\\%
&$\Delta$&8.09\%&16.86\%&18.20\%&14.31\%&14.21\%&{-}47.33\%&{-}113.31\%&{-}64.37\%&{-}45.26\%&{-}64.39\%&\\%
\hline%
\multirow{1}{*}{3d\_supervised}&N/A&89.01&91.68&81.44&70.30&83.11&4.73&5.14&5.91&6.83&5.65&\multirow{1}{*}{5.75}\\%
\hline%
\end{tabular}}
\begin{table*}[t!]
    \centering
    \color{black}
    \caption{\label{tab:Dicethree} Dice score (\%) and 95\% Hausdorff distance achieved when institution 3 was adopted as the novel institution. `support ins' refers to the institution from which the support images were sampled from. $\Delta$ refers to the percentage difference between predictions made with support images from the base and novel institutions.}
    \small
\begin{tabular}{|c|c|ccccc|ccccc|c|}%
\hline%
\multirow{2}{*}{model}&support&\multicolumn{5}{c|}{Dice (\%)}&\multicolumn{5}{c|}{95\% Hausdorff distance (mm)}&\#param\\%
&ins&fold1&fold2&fold3&fold4&mean&fold1&fold2&fold3&fold4&mean&(million)\\%
\hline%
\multirow{4}{*}{3d\_finetune}&all&48.07&44.78&43.92&29.25&41.50&61.47&69.78&60.33&57.02&62.15&\multirow{4}{*}{5.75}\\%
&base&45.74&39.81&41.69&26.02&38.32&63.22&73.42&62.33&59.24&64.55&\\%
&novel&62.03&74.56&57.31&48.63&60.63&50.98&47.96&48.32&43.72&47.74&\\%
&$\Delta$&26.26\%&46.60\%&27.25\%&46.48\%&36.80\%&{-}24.02\%&{-}53.09\%&{-}28.99\%&{-}35.51\%&{-}35.21\%&\\%
\hline%
\multirow{4}{*}{BiGRU}&all&44.05&43.08&29.61&20.97&34.43&23.70&33.67&23.60&22.83&25.95&\multirow{4}{*}{38.85}\\%
&base&44.05&42.74&29.21&20.62&34.15&23.73&34.06&23.85&23.20&26.21&\\%
&novel&44.03&45.11&32.01&23.12&36.07&23.49&31.30&22.12&20.60&24.38&\\%
&$\Delta$&{-}0.06\%&5.26\%&8.75\%&10.80\%&5.30\%&{-}1.04\%&{-}8.82\%&{-}7.83\%&{-}12.59\%&{-}7.52\%&\\
\hline%
\multirow{4}{*}{2d}&all&49.23&44.51&36.34&27.44&39.38&24.03&30.86&29.51&30.77&28.79&\multirow{4}{*}{23.63}\\%
&base&48.80&42.92&35.52&26.83&38.52&24.12&31.93&30.30&31.54&29.47&\\%
&novel&51.80&54.07&41.28&31.08&44.56&23.52&24.39&24.83&26.13&24.72&\\%
&$\Delta$&5.80\%&20.62\%&13.94\%&13.68\%&13.55\%&{-}2.54\%&{-}30.92\%&{-}22.03\%&{-}20.72\%&{-}19.24\%&\\%
\hline%
\multirow{4}{*}{3d}&all&47.79&52.06&34.81&26.16&40.20&31.60&30.68&32.95&31.71&31.73&\multirow{4}{*}{5.75}\\%
&base&47.89&51.07&34.12&25.44&39.63&31.69&31.03&33.35&32.19&32.07&\\%
&novel&47.22&58.00&38.96&30.49&43.67&31.07&28.54&30.54&28.81&29.74&\\%
&$\Delta$&{-}1.42\%&11.94\%&12.44\%&16.58\%&9.25\%&{-}2.00\%&{-}8.73\%&{-}9.20\%&{-}11.73\%&{-}7.82\%&\\%
\hline%
\multirow{4}{*}{3d\_con}&all&58.34&46.11&40.43&30.94&43.96&27.31&27.15&27.33&26.38&27.04&\multirow{4}{*}{5.75}\\%
&base&58.25&43.58&39.87&29.80&42.88&27.69&28.34&28.03&27.06&27.78&\\%
&novel&58.84&61.30&43.81&37.78&50.43&25.04&20.02&23.14&22.30&22.63&\\%
&$\Delta$&1.00\%&28.91\%&8.99\%&21.12\%&14.98\%&{-}10.56\%&{-}41.54\%&{-}21.12\%&{-}21.31\%&{-}22.76\%&\\%
\hline%
\multirow{4}{*}{3d\_align}&all&50.80&51.56&32.39&30.89&41.41&27.33&36.29&32.63&29.93&31.55&\multirow{4}{*}{27.34}\\%
&base&50.60&51.17&31.25&30.82&40.96&27.35&36.82&33.01&30.42&31.90&\\%
&novel&51.97&53.88&39.22&31.33&44.10&27.20&33.15&30.34&26.96&29.41&\\%
&$\Delta$&2.64\%&5.04\%&20.31\%&1.62\%&7.12\%&{-}0.55\%&{-}11.08\%&{-}8.78\%&{-}12.84\%&{-}8.46\%&\\%
\hline%
\multirow{4}{*}{3d\_con\_align}&all&59.36&60.38&45.73&37.60&50.77&22.66&29.63&27.92&26.41&26.66&\multirow{4}{*}{27.35}\\%
&base&59.11&59.80&45.06&37.28&50.31&22.69&30.30&28.38&26.89&27.06&\\%
&novel&60.86&63.87&49.77&39.54&53.51&22.49&25.63&25.19&23.54&24.21&\\%
&$\Delta$&2.89\%&6.37\%&9.47\%&5.70\%&5.98\%&{-}0.91\%&{-}18.21\%&{-}12.65\%&{-}14.26\%&{-}11.79\%&\\%
\hline%
\multirow{2}{*}{localnet}&all&25.64&38.99&21.83&11.20&24.42&28.74&27.21&24.78&26.97&26.93&\multirow{4}{*}{5.94}\\%
&base&24.16&36.43&21.03&10.35&22.99&29.43&28.48&25.39&27.89&27.80&\\%
\multirow{2}{*}{(unsupervised)}&novel&34.52&54.37&26.61&16.34&32.96&24.60&19.55&21.10&21.47&21.68&\\%
&$\Delta$&30.01\%&33.00\%&20.98\%&36.67\%&30.25\%&{-}19.63\%&{-}45.70\%&{-}20.32\%&{-}29.93\%&{-}28.22\%&\\%
\hline%
\multirow{2}{*}{localnet}&all&72.70&72.14&52.84&42.65&60.08&19.36&22.81&21.78&24.00&21.99&\multirow{4}{*}{5.94}\\%
&base&71.80&70.11&51.21&41.65&58.69&20.29&24.69&23.07&25.11&23.29&\\%
\multirow{2}{*}{(supervised)}&novel&78.11&84.33&62.61&48.61&68.42&13.77&11.57&14.03&17.29&14.17&\\%
&$\Delta$&8.09\%&16.86\%&18.20\%&14.31\%&14.21\%&{-}47.33\%&{-}113.31\%&{-}64.37\%&{-}45.26\%&{-}64.39\%&\\%
\hline%
\multirow{1}{*}{3d\_supervised}&N/A&89.01&91.68&81.44&70.30&83.11&4.73&5.14&5.91&6.83&5.65&\multirow{1}{*}{5.75}\\%
\hline%
\end{tabular}
\end{table*}

\newcommand{\Dicefour}{\small
\begin{tabular}{|c|c|ccccc|ccccc|}%
\hline%
\multirow{2}{*}{model}&support&\multicolumn{5}{c|}{Dice (\%)}&\multicolumn{5}{c|}{95\% Hausdorff distance (mm)}\\%
&ins&fold1&fold2&fold3&fold4&mean&fold1&fold2&fold3&fold4&mean\\%
\hline%
\multirow{4}{*}{3d\_finetune}&all&37.61&29.23&33.06&20.25&30.04&59.83&71.87&50.66&51.62&58.49\\%
&base&36.72&28.33&31.11&20.02&29.05&60.77&73.85&52.09&52.02&59.68\\%
&novel&42.96&34.63&44.73&21.63&35.99&54.16&59.97&42.09&49.20&51.36\\%
&$\Delta$&14.53\%&18.17\%&30.45\%&7.43\%&19.29\%&{-}12.21\%&{-}23.13\%&{-}23.75\%&{-}5.74\%&{-}16.22\%\\%
\hline%
\multirow{4}{*}{2d}&all&42.09&29.00&32.49&24.46&32.01&29.52&42.65&28.04&31.97&33.05\\%
&base&41.83&28.51&32.09&24.47&31.73&29.62&42.39&28.08&31.93&33.01\\%
&novel&43.64&31.92&34.85&24.41&33.71&28.90&44.18&27.80&32.19&33.27\\%
&$\Delta$&4.14\%&10.68\%&7.92\%&{-}0.26\%&5.87\%&{-}2.50\%&4.05\%&{-}1.00\%&0.80\%&0.79\%\\%
\hline%
\multirow{4}{*}{3d\_con\_align}&all&57.58&37.42&37.42&30.18&40.65&23.02&37.56&30.22&28.52&29.83\\%
&base&57.71&36.74&36.70&29.97&40.28&22.90&38.37&30.74&28.52&30.13\\%
&novel&56.83&41.46&41.70&31.40&42.85&23.75&32.72&27.12&28.51&28.02\\%
&$\Delta$&{-}1.54\%&11.39\%&11.99\%&4.56\%&6.00\%&3.58\%&{-}17.26\%&{-}13.36\%&{-}0.04\%&{-}7.52\%\\%
\hline%
\multirow{1}{*}{3d\_supervised}&N/A&88.31&88.44&80.49&70.78&82.00&5.14&7.72&6.49&6.71&6.51\\%
\hline%
\end{tabular}}
\begin{table*}[t!]
    \centering
    \caption{\label{tab:Dicefour} Dice score (\%) and 95\% Hausdorff distance achieved when institution 4 was adopted as the novel institution. `support ins' refers to the institution from which the support images were sampled from. $\Delta$ refers to the percentage difference between predictions made with support images from the base and novel institutions.}
    \small
\begin{tabular}{|c|c|ccccc|ccccc|}%
\hline%
\multirow{2}{*}{model}&support&\multicolumn{5}{c|}{Dice (\%)}&\multicolumn{5}{c|}{95\% Hausdorff distance (mm)}\\%
&ins&fold1&fold2&fold3&fold4&mean&fold1&fold2&fold3&fold4&mean\\%
\hline%
\multirow{4}{*}{3d\_finetune}&all&37.61&29.23&33.06&20.25&30.04&59.83&71.87&50.66&51.62&58.49\\%
&base&36.72&28.33&31.11&20.02&29.05&60.77&73.85&52.09&52.02&59.68\\%
&novel&42.96&34.63&44.73&21.63&35.99&54.16&59.97&42.09&49.20&51.36\\%
&$\Delta$&14.53\%&18.17\%&30.45\%&7.43\%&19.29\%&{-}12.21\%&{-}23.13\%&{-}23.75\%&{-}5.74\%&{-}16.22\%\\%
\hline%
\multirow{4}{*}{2d}&all&42.09&29.00&32.49&24.46&32.01&29.52&42.65&28.04&31.97&33.05\\%
&base&41.83&28.51&32.09&24.47&31.73&29.62&42.39&28.08&31.93&33.01\\%
&novel&43.64&31.92&34.85&24.41&33.71&28.90&44.18&27.80&32.19&33.27\\%
&$\Delta$&4.14\%&10.68\%&7.92\%&{-}0.26\%&5.87\%&{-}2.50\%&4.05\%&{-}1.00\%&0.80\%&0.79\%\\%
\hline%
\multirow{4}{*}{3d\_con\_align}&all&57.58&37.42&37.42&30.18&40.65&23.02&37.56&30.22&28.52&29.83\\%
&base&57.71&36.74&36.70&29.97&40.28&22.90&38.37&30.74&28.52&30.13\\%
&novel&56.83&41.46&41.70&31.40&42.85&23.75&32.72&27.12&28.51&28.02\\%
&$\Delta$&{-}1.54\%&11.39\%&11.99\%&4.56\%&6.00\%&3.58\%&{-}17.26\%&{-}13.36\%&{-}0.04\%&{-}7.52\%\\%
\hline%
\multirow{1}{*}{3d\_supervised}&N/A&88.31&88.44&80.49&70.78&82.00&5.14&7.72&6.49&6.71&6.51\\%
\hline%
\end{tabular}
\end{table*}

\newcommand{\BaseIns}{\begin{tabular}{|c|c|ccccccc|c|}%
\hline%
\multirow{3}{*}{model}&support&\multicolumn{7}{c|}{Dice(\%)}&\#param\\%
&ins&\multicolumn{7}{c|}{query institution}&(million)\\%
&&ins1&ins2&ins4&ins5&ins6&ins7&mean&\\%
\hline%
\multirow{4}{*}{BiGRU}&same&41.67&45.03&34.16&38.25&36.63&27.43&37.20&\multirow{4}{*}{38.85}\\%
&diff&35.40&34.00&31.60&34.09&36.77&25.76&32.94&\\%
&all&36.18&35.38&31.92&34.61&36.75&25.97&33.47&\\%
&$\Delta$&15.06\%&24.49\%&7.49\%&10.88\%&{-}0.37\%&6.09\%&11.45\%&\\%
\hline%
\multirow{4}{*}{2d}&same&48.41&49.15&39.00&53.49&44.33&34.52&44.82&\multirow{4}{*}{23.63}\\%
&diff&39.14&37.89&33.15&38.74&38.83&30.94&36.45&\\%
&all&40.30&39.30&33.88&40.58&39.51&31.38&37.49&\\%
&$\Delta$&19.15\%&22.91\%&15.02\%&27.58\%&12.42\%&10.38\%&18.68\%&\\%
\hline%
\multirow{4}{*}{3d\_con\_align}&same&53.76&57.74&50.73&61.14&52.02&49.85&54.21&\multirow{4}{*}{27.35}\\%
&diff&49.55&50.39&47.64&53.73&51.38&46.22&49.82&\\%
&all&50.08&51.30&48.03&54.65&51.46&46.67&50.37&\\%
&$\Delta$&7.83\%&12.73\%&6.08\%&12.12\%&1.23\%&7.29\%&8.10\%&\\%
\hline%
\multirow{2}{*}{localnet}&same&36.42&48.05&27.66&39.53&26.51&19.94&33.02&\multirow{4}{*}{5.94}\\%
&diff&27.50&30.88&21.77&20.28&29.00&20.88&25.05&\\%
\multirow{2}{*}{(unsupervised)}&all&28.62&33.03&22.50&22.69&28.69&20.76&26.05&\\%
&$\Delta$&24.49\%&35.73\%&21.30\%&48.69\%&{-}9.37\%&{-}4.68\%&24.13\%&\\%
\hline%
\multirow{2}{*}{localnet}&same&73.05&64.65&41.41&73.38&68.87&40.44&60.30&\multirow{4}{*}{5.94}\\%
&diff&62.48&58.76&50.72&61.92&60.79&51.28&57.66&\\%
\multirow{2}{*}{(supervised)}&all&63.80&59.50&49.56&63.35&61.80&49.92&57.99&\\%
&$\Delta$&14.47\%&9.11\%&{-}22.49\%&15.62\%&11.74\%&{-}26.81\%&4.38\%&\\%
\hline%
\multirow{1}{*}{3d\_supervised}&N/A&86.77&84.08&80.75&86.36&80.16&80.69&83.14&\multirow{1}{*}{5.75}\\%
\hline%
\end{tabular}}
\begin{table*}[h]
    \color{black}
    \centering
    \caption{\label{tab:BaseIns} Dice score (\%) achieved when query and support both came from base institutions while institution 3 was adopted as the novel institution. `support ins' here denote if the support and query come from the same institution, different institutions or all base institutions are counted.}
    \begin{tabular}{|c|c|ccccccc|c|}%
\hline%
\multirow{3}{*}{model}&support&\multicolumn{7}{c|}{Dice(\%)}&\#param\\%
&ins&\multicolumn{7}{c|}{query institution}&(million)\\%
&&ins1&ins2&ins4&ins5&ins6&ins7&mean&\\%
\hline%
\multirow{4}{*}{BiGRU}&same&41.67&45.03&34.16&38.25&36.63&27.43&37.20&\multirow{4}{*}{38.85}\\%
&diff&35.40&34.00&31.60&34.09&36.77&25.76&32.94&\\%
&all&36.18&35.38&31.92&34.61&36.75&25.97&33.47&\\%
&$\Delta$&15.06\%&24.49\%&7.49\%&10.88\%&{-}0.37\%&6.09\%&11.45\%&\\%
\hline%
\multirow{4}{*}{2d}&same&48.41&49.15&39.00&53.49&44.33&34.52&44.82&\multirow{4}{*}{23.63}\\%
&diff&39.14&37.89&33.15&38.74&38.83&30.94&36.45&\\%
&all&40.30&39.30&33.88&40.58&39.51&31.38&37.49&\\%
&$\Delta$&19.15\%&22.91\%&15.02\%&27.58\%&12.42\%&10.38\%&18.68\%&\\%
\hline%
\multirow{4}{*}{3d\_con\_align}&same&53.76&57.74&50.73&61.14&52.02&49.85&54.21&\multirow{4}{*}{27.35}\\%
&diff&49.55&50.39&47.64&53.73&51.38&46.22&49.82&\\%
&all&50.08&51.30&48.03&54.65&51.46&46.67&50.37&\\%
&$\Delta$&7.83\%&12.73\%&6.08\%&12.12\%&1.23\%&7.29\%&8.10\%&\\%
\hline%
\multirow{2}{*}{localnet}&same&36.42&48.05&27.66&39.53&26.51&19.94&33.02&\multirow{4}{*}{5.94}\\%
&diff&27.50&30.88&21.77&20.28&29.00&20.88&25.05&\\%
\multirow{2}{*}{(unsupervised)}&all&28.62&33.03&22.50&22.69&28.69&20.76&26.05&\\%
&$\Delta$&24.49\%&35.73\%&21.30\%&48.69\%&{-}9.37\%&{-}4.68\%&24.13\%&\\%
\hline%
\multirow{2}{*}{localnet}&same&73.05&64.65&41.41&73.38&68.87&40.44&60.30&\multirow{4}{*}{5.94}\\%
&diff&62.48&58.76&50.72&61.92&60.79&51.28&57.66&\\%
\multirow{2}{*}{(supervised)}&all&63.80&59.50&49.56&63.35&61.80&49.92&57.99&\\%
&$\Delta$&14.47\%&9.11\%&{-}22.49\%&15.62\%&11.74\%&{-}26.81\%&4.38\%&\\%
\hline%
\multirow{1}{*}{3d\_supervised}&N/A&86.77&84.08&80.75&86.36&80.16&80.69&83.14&\multirow{1}{*}{5.75}\\%
\hline%
\end{tabular}
\end{table*}

\CatchFileDef{\cross2d}{baseline_2d_cross_ins.tex}{}
\begin{table*}[t!]
    \centering
    \caption{\label{tab:cross_2d} Mean Dice score (\%) achieved at different (s\_ins, q\_ins) combinations by `2d' when Institution 3 was adopted as the novel institution, where
    s\_ins and q\_ins respectively refers to the institution from which the support and query were sampled from. Best performance for each query institution were bolded. p-values were derived from paired t-test performed between the Dice scores where support institution equals query institution and the maximum Dice scores achieved when support institution differs from the query institution.}
    \cross2d
\end{table*}

\newcommand{\crossours}{\small
\begin{tabular}{|cc|ccccccc|ccc|}%
\hline%
&&\multicolumn{7}{c|}{s\_ins}&std&mean&p{-}value\\%
&&ins1&ins2&ins3&ins4&ins5&ins6&ins7&&&\\%
\hline%
\multirow{7}{*}{q\_ins}&ins1&\textbf{53.76}&53.31&50.63&46.46&50.77&47.78&47.83&2.62&50.08& 1.06e{-}15\\%
&ins2&54.26&\textbf{57.74}&50.07&44.42&51.97&52.05&48.62&3.91&51.30& 0.08\\%
&ins3&51.11&50.17&\textbf{53.51}&49.41&52.25&47.25&51.67&1.90&50.77& 2.04e{-}16\\%
&ins4&48.23&44.70&49.95&50.73&47.09&44.02&\textbf{51.47}&2.70&48.03& 1.40e{-}05\\%
&ins5&54.70&52.88&54.38&54.14&\textbf{61.14}&50.56&54.77&2.98&54.65& 0.15\\%
&ins6&52.10&\textbf{55.89}&51.69&44.23&52.70&52.02&51.61&3.26&51.46& 9.47e{-}03\\%
&ins7&46.46&45.81&48.77&48.78&45.45&41.57&\textbf{49.85}&2.60&46.67& 0.20\\%
\hline%
\end{tabular}}
\begin{table*}[t!]
    \centering
    \caption{\label{tab:cross_ours} Mean Dice score (\%) achieved at different (s\_ins, q\_ins) combinations by `3d\_con\_align' when Institution 3 was adopted as the novel institution, where
    s\_ins and q\_ins respectively refers to the institution from which the support and query were sampled from. Best performance for each query institution were bolded. p-values were derived from paired t-test performed between the Dice scores where support institution equals query institution and the maximum Dice scores achieved when support institution differs from the query institution.}
    \small
\begin{tabular}{|cc|ccccccc|ccc|}%
\hline%
&&\multicolumn{7}{c|}{s\_ins}&std&mean&p{-}value\\%
&&ins1&ins2&ins3&ins4&ins5&ins6&ins7&&&\\%
\hline%
\multirow{7}{*}{q\_ins}&ins1&\textbf{53.76}&53.31&50.63&46.46&50.77&47.78&47.83&2.62&50.08& 1.06e{-}15\\%
&ins2&54.26&\textbf{57.74}&50.07&44.42&51.97&52.05&48.62&3.91&51.30& 0.08\\%
&ins3&51.11&50.17&\textbf{53.51}&49.41&52.25&47.25&51.67&1.90&50.77& 2.04e{-}16\\%
&ins4&48.23&44.70&49.95&50.73&47.09&44.02&\textbf{51.47}&2.70&48.03& 1.40e{-}05\\%
&ins5&54.70&52.88&54.38&54.14&\textbf{61.14}&50.56&54.77&2.98&54.65& 0.15\\%
&ins6&52.10&\textbf{55.89}&51.69&44.23&52.70&52.02&51.61&3.26&51.46& 9.47e{-}03\\%
&ins7&46.46&45.81&48.77&48.78&45.45&41.57&\textbf{49.85}&2.60&46.67& 0.20\\%
\hline%
\end{tabular}
\end{table*}

\newcommand{\trainingsizeablation}{\small
\begin{tabular}{|c|c|ccccc|ccccc|}%
\hline%
training&\multirow{2}{*}{s\_ins}&\multicolumn{5}{c|}{Dice (\%)}&\multicolumn{5}{c|}{95\% Hausdorff distance (mm)}\\%
data&&fold1&fold2&fold3&fold4&mean&fold1&fold2&fold3&fold4&mean\\%
\hline%
\multirow{4}{*}{whole}&all&59.36&60.38&45.73&37.60&50.77&22.66&29.63&27.92&26.41&26.66\\%
&base&59.11&59.80&45.06&37.28&50.31&22.69&30.30&28.38&26.89&27.06\\%
&novel&60.86&63.87&49.77&39.54&53.51&22.49&25.63&25.19&23.54&24.21\\%
&$\Delta$&2.89\%&6.37\%&9.47\%&5.70\%&5.98\%&{-}0.91\%&{-}18.21\%&{-}12.65\%&{-}14.26\%&{-}11.79\%\\%
\hline%
\multirow{4}{*}{half}&all&55.38&57.16&42.54&34.14&47.30&25.98&29.37&31.65&29.00&29.00\\%
&base&55.08&55.49&41.82&33.75&46.54&26.04&30.13&32.17&29.37&29.43\\%
&novel&57.16&67.17&46.88&36.46&51.91&25.56&24.81&28.52&26.84&26.43\\%
&$\Delta$&3.63\%&17.38\%&10.79\%&7.41\%&10.36\%&{-}1.91\%&{-}21.45\%&{-}12.82\%&{-}9.43\%&{-}11.35\%\\%
\hline%
\multirow{4}{*}{half\_single\_ins}&all&52.85&57.46&31.31&20.39&40.50&26.84&29.38&35.41&33.89&31.38\\%
&base&52.17&56.06&30.55&19.63&39.60&26.93&29.93&36.04&34.41&31.83\\%
&novel&56.90&65.87&35.85&24.93&45.89&26.29&26.10&31.65&30.75&28.70\\%
&$\Delta$&8.32\%&14.89\%&14.79\%&21.24\%&13.69\%&{-}2.44\%&{-}14.63\%&{-}13.88\%&{-}11.89\%&{-}10.90\%\\%
\hline%
\multirow{4}{*}{quarter}&all&50.90&55.87&44.78&31.98&45.88&27.90&29.61&28.87&29.48&28.96\\%
&base&50.35&54.54&43.64&31.61&45.03&27.90&30.22&29.43&29.84&29.35\\%
&novel&54.24&63.86&51.61&34.17&50.97&27.88&25.92&25.50&27.27&26.64\\%
&$\Delta$&7.17\%&14.60\%&15.45\%&7.49\%&11.65\%&{-}0.05\%&{-}16.60\%&{-}15.41\%&{-}9.43\%&{-}10.15\%\\%
\hline%
\end{tabular}}
\begin{table*}[t!]
    \centering
    \caption{\label{tab:trainsize}Dice score (\%) and 95\% Hausdorff distance achieved by `3d\_con\_align' with various training data availability when Institution 3 was adopted as the novel institution. `support ins' refers to the institution from which the support was sampled from. $\Delta$ refers to the percentage difference when the support image comes from the base and novel institutions. }
    \small
\begin{tabular}{|c|c|ccccc|ccccc|}%
\hline%
training&\multirow{2}{*}{s\_ins}&\multicolumn{5}{c|}{Dice (\%)}&\multicolumn{5}{c|}{95\% Hausdorff distance (mm)}\\%
data&&fold1&fold2&fold3&fold4&mean&fold1&fold2&fold3&fold4&mean\\%
\hline%
\multirow{4}{*}{whole}&all&59.36&60.38&45.73&37.60&50.77&22.66&29.63&27.92&26.41&26.66\\%
&base&59.11&59.80&45.06&37.28&50.31&22.69&30.30&28.38&26.89&27.06\\%
&novel&60.86&63.87&49.77&39.54&53.51&22.49&25.63&25.19&23.54&24.21\\%
&$\Delta$&2.89\%&6.37\%&9.47\%&5.70\%&5.98\%&{-}0.91\%&{-}18.21\%&{-}12.65\%&{-}14.26\%&{-}11.79\%\\%
\hline%
\multirow{4}{*}{half}&all&55.38&57.16&42.54&34.14&47.30&25.98&29.37&31.65&29.00&29.00\\%
&base&55.08&55.49&41.82&33.75&46.54&26.04&30.13&32.17&29.37&29.43\\%
&novel&57.16&67.17&46.88&36.46&51.91&25.56&24.81&28.52&26.84&26.43\\%
&$\Delta$&3.63\%&17.38\%&10.79\%&7.41\%&10.36\%&{-}1.91\%&{-}21.45\%&{-}12.82\%&{-}9.43\%&{-}11.35\%\\%
\hline%
\multirow{4}{*}{half\_single\_ins}&all&52.85&57.46&31.31&20.39&40.50&26.84&29.38&35.41&33.89&31.38\\%
&base&52.17&56.06&30.55&19.63&39.60&26.93&29.93&36.04&34.41&31.83\\%
&novel&56.90&65.87&35.85&24.93&45.89&26.29&26.10&31.65&30.75&28.70\\%
&$\Delta$&8.32\%&14.89\%&14.79\%&21.24\%&13.69\%&{-}2.44\%&{-}14.63\%&{-}13.88\%&{-}11.89\%&{-}10.90\%\\%
\hline%
\multirow{4}{*}{quarter}&all&50.90&55.87&44.78&31.98&45.88&27.90&29.61&28.87&29.48&28.96\\%
&base&50.35&54.54&43.64&31.61&45.03&27.90&30.22&29.43&29.84&29.35\\%
&novel&54.24&63.86&51.61&34.17&50.97&27.88&25.92&25.50&27.27&26.64\\%
&$\Delta$&7.17\%&14.60\%&15.45\%&7.49\%&11.65\%&{-}0.05\%&{-}16.60\%&{-}15.41\%&{-}9.43\%&{-}10.15\%\\%
\hline%
\end{tabular}
\end{table*}

\newcommand{\kshotablation}{\small
\begin{tabular}{|c|c|c|ccccc|ccccc|}%
\hline%
\multirow{2}{*}{shot}&\multirow{2}{*}{\# of shot}&support&\multicolumn{5}{c|}{Dice (\%)}&\multicolumn{5}{c|}{95\% Hausdorff distance (mm)}\\%
&&ins&fold1&fold2&fold3&fold4&mean&fold1&fold2&fold3&fold4&mean\\%
\hline%
\multirow{12}{*}{1}&\multirow{4}{*}{3d\_finetune}&all&48.07&44.78&43.92&29.25&41.50&61.47&69.78&60.33&57.02&62.15\\%
&&base&45.74&39.81&41.69&26.02&38.32&63.22&73.42&62.33&59.24&64.55\\%
&&novel&62.03&74.56&57.31&48.63&60.63&50.98&47.96&48.32&43.72&47.74\\%
&&$\Delta$&26.26\%&46.60\%&27.25\%&46.48\%&36.80\%&{-}24.02\%&{-}53.09\%&{-}28.99\%&{-}35.51\%&{-}35.21\%\\%
\cline{2%
-%
13}%
&\multirow{4}{*}{3d\_con\_align}&all&59.36&60.38&45.73&37.60&50.77&22.66&29.63&27.92&26.41&26.66\\%
&&base&59.11&59.80&45.06&37.28&50.31&22.69&30.30&28.38&26.89&27.06\\%
&&novel&60.86&63.87&49.77&39.54&53.51&22.49&25.63&25.19&23.54&24.21\\%
&&$\Delta$&2.89\%&6.37\%&9.47\%&5.70\%&5.98\%&{-}0.91\%&{-}18.21\%&{-}12.65\%&{-}14.26\%&{-}11.79\%\\%
\cline{2%
-%
13}%
&\multirow{2}{*}{localnet}&all&72.70&72.14&52.84&42.65&60.08&19.36&22.81&21.78&24.00&21.99\\%
&&base&71.80&70.11&51.21&41.65&58.69&20.29&24.69&23.07&25.11&23.29\\%
&\multirow{2}{*}{(supervised)}&novel&78.11&84.33&62.61&48.61&68.42&13.77&11.57&14.03&17.29&14.17\\%
&&$\Delta$&8.09\%&16.86\%&18.20\%&14.31\%&14.21\%&{-}47.33\%&{-}113.31\%&{-}64.37\%&{-}45.26\%&{-}64.39\%\\%
\hline%
\multirow{12}{*}{2}&\multirow{4}{*}{3d\_finetune}&all&52.58&57.19&45.20&35.41&47.59&52.90&55.99&44.63&49.53&50.77\\%
&&base&49.94&52.74&43.50&32.35&44.63&55.69&60.55&47.35&53.64&54.31\\%
&&novel&68.42&83.86&55.43&53.81&65.38&36.18&28.60&28.30&24.88&29.49\\%
&&$\Delta$&27.01\%&37.11\%&21.53\%&39.89\%&31.74\%&{-}53.91\%&{-}111.71\%&{-}67.33\%&{-}115.65\%&{-}84.16\%\\%
\cline{2%
-%
13}%
&\multirow{4}{*}{3d\_con\_align}&all&60.64&60.41&47.82&38.91&51.95&21.63&29.19&27.08&25.66&25.89\\%
&&base&60.03&59.82&47.02&38.54&51.35&21.80&29.89&27.55&26.07&26.33\\%
&&novel&64.28&63.94&52.66&41.12&55.50&20.59&24.99&24.27&23.22&23.27\\%
&&$\Delta$&6.61\%&6.44\%&10.71\%&6.27\%&7.47\%&{-}5.88\%&{-}19.59\%&{-}13.52\%&{-}12.26\%&{-}13.15\%\\%
\cline{2%
-%
13}%
&\multirow{2}{*}{localnet}&all&71.93&71.47&50.99&43.23&59.40&10.93&21.92&15.91&16.36&16.28\\%
&&base&70.64&68.98&48.68&41.90&57.55&11.42&24.53&16.81&17.05&17.45\\%
&\multirow{2}{*}{(supervised)}&novel&79.66&86.41&64.84&51.21&70.53&7.99&6.25&10.48&12.23&9.24\\%
&&$\Delta$&11.32\%&20.18\%&24.92\%&18.19\%&18.41\%&{-}42.99\%&{-}292.69\%&{-}60.36\%&{-}39.35\%&{-}88.93\%\\%
\hline%
\multirow{12}{*}{3}&\multirow{4}{*}{3d\_finetune}&all&53.92&60.16&47.94&36.24&49.57&52.14&52.74&40.64&49.66&48.80\\%
&&base&51.61&56.12&46.51&33.30&46.88&54.27&57.69&43.23&52.82&52.00\\%
&&novel&67.80&84.38&56.56&53.90&65.66&39.34&23.04&25.12&30.69&29.54\\%
&&$\Delta$&23.88\%&33.50\%&17.77\%&38.23\%&28.60\%&{-}37.96\%&{-}150.45\%&{-}72.11\%&{-}72.12\%&{-}76.02\%\\%
\cline{2%
-%
13}%
&\multirow{4}{*}{3d\_con\_align}&all&61.36&60.93&48.53&38.55&52.34&21.59&28.81&26.66&25.66&25.68\\%
&&base&61.06&60.27&47.99&38.25&51.89&21.64&29.42&26.87&26.06&26.00\\%
&&novel&63.13&64.88&51.80&40.33&55.03&21.33&25.15&25.42&23.28&23.80\\%
&&$\Delta$&3.28\%&7.10\%&7.35\%&5.16\%&5.71\%&{-}1.43\%&{-}16.96\%&{-}5.68\%&{-}11.93\%&{-}9.24\%\\%
\cline{2%
-%
13}%
&\multirow{2}{*}{localnet}&all&78.07&74.86&56.98&47.94&64.46&12.73&16.10&16.97&19.12&16.23\\%
&&base&77.57&73.08&55.73&47.09&63.37&13.50&17.41&17.89&19.98&17.19\\%
&\multirow{2}{*}{(supervised)}&novel&81.09&85.53&64.48&53.06&71.04&8.11&8.23&11.46&13.98&10.45\\%
&&$\Delta$&4.35\%&14.55\%&13.57\%&11.24\%&10.80\%&{-}66.36\%&{-}111.55\%&{-}56.08\%&{-}42.92\%&{-}64.60\%\\%
\hline%
\multirow{12}{*}{4}&\multirow{4}{*}{3d\_finetune}&all&55.53&61.34&48.17&36.85&50.47&50.70&49.19&39.65&47.11&46.66\\%
&&base&52.93&56.80&46.32&33.40&47.36&53.41&54.70&43.12&51.48&50.68\\%
&&novel&68.50&84.07&57.37&54.06&66.00&37.17&21.62&22.29&25.30&26.60\\%
&&$\Delta$&22.73\%&32.44\%&19.25\%&38.21\%&28.24\%&{-}43.66\%&{-}153.05\%&{-}93.42\%&{-}103.48\%&{-}90.54\%\\%
\cline{2%
-%
13}%
&\multirow{4}{*}{3d\_con\_align}&all&61.33&60.63&49.52&38.16&52.41&21.46&29.59&26.14&26.32&25.88\\%
&&base&60.69&59.91&49.01&37.56&51.79&21.64&30.34&26.64&27.06&26.42\\%
&&novel&64.52&64.23&52.08&41.19&55.50&20.57&25.80&23.64&22.63&23.16\\%
&&$\Delta$&5.94\%&6.73\%&5.90\%&8.81\%&6.69\%&{-}5.19\%&{-}17.63\%&{-}12.69\%&{-}19.57\%&{-}14.08\%\\%
\cline{2%
-%
13}%
&\multirow{2}{*}{localnet}&all&77.64&81.89&60.06&51.96&67.89&8.95&9.62&12.77&13.38&11.18\\%
&&base&76.73&81.04&58.84&51.44&67.01&9.50&10.28&13.32&13.75&11.71\\%
&\multirow{2}{*}{(supervised)}&novel&82.14&86.15&66.18&54.57&72.26&6.23&6.33&10.02&11.54&8.53\\%
&&$\Delta$&6.58\%&5.94\%&11.09\%&5.73\%&7.26\%&{-}52.50\%&{-}62.41\%&{-}32.94\%&{-}19.22\%&{-}37.34\%\\%
\hline%
\end{tabular}}
\begin{table*}[t!]
    \centering
    \color{black}
    \caption{Dice score achieved by `3d\_con\_align' with various number of shots when Institution 3 was adopted as the novel institution. The `support ins' refers to the institution from which the support images were sampled from. }
    \small
\begin{tabular}{|c|c|c|ccccc|ccccc|}%
\hline%
\multirow{2}{*}{shot}&\multirow{2}{*}{\# of shot}&support&\multicolumn{5}{c|}{Dice (\%)}&\multicolumn{5}{c|}{95\% Hausdorff distance (mm)}\\%
&&ins&fold1&fold2&fold3&fold4&mean&fold1&fold2&fold3&fold4&mean\\%
\hline%
\multirow{12}{*}{1}&\multirow{4}{*}{3d\_finetune}&all&48.07&44.78&43.92&29.25&41.50&61.47&69.78&60.33&57.02&62.15\\%
&&base&45.74&39.81&41.69&26.02&38.32&63.22&73.42&62.33&59.24&64.55\\%
&&novel&62.03&74.56&57.31&48.63&60.63&50.98&47.96&48.32&43.72&47.74\\%
&&$\Delta$&26.26\%&46.60\%&27.25\%&46.48\%&36.80\%&{-}24.02\%&{-}53.09\%&{-}28.99\%&{-}35.51\%&{-}35.21\%\\%
\cline{2%
-%
13}%
&\multirow{4}{*}{3d\_con\_align}&all&59.36&60.38&45.73&37.60&50.77&22.66&29.63&27.92&26.41&26.66\\%
&&base&59.11&59.80&45.06&37.28&50.31&22.69&30.30&28.38&26.89&27.06\\%
&&novel&60.86&63.87&49.77&39.54&53.51&22.49&25.63&25.19&23.54&24.21\\%
&&$\Delta$&2.89\%&6.37\%&9.47\%&5.70\%&5.98\%&{-}0.91\%&{-}18.21\%&{-}12.65\%&{-}14.26\%&{-}11.79\%\\%
\cline{2%
-%
13}%
&\multirow{2}{*}{localnet}&all&72.70&72.14&52.84&42.65&60.08&19.36&22.81&21.78&24.00&21.99\\%
&&base&71.80&70.11&51.21&41.65&58.69&20.29&24.69&23.07&25.11&23.29\\%
&\multirow{2}{*}{(supervised)}&novel&78.11&84.33&62.61&48.61&68.42&13.77&11.57&14.03&17.29&14.17\\%
&&$\Delta$&8.09\%&16.86\%&18.20\%&14.31\%&14.21\%&{-}47.33\%&{-}113.31\%&{-}64.37\%&{-}45.26\%&{-}64.39\%\\%
\hline%
\multirow{12}{*}{2}&\multirow{4}{*}{3d\_finetune}&all&52.58&57.19&45.20&35.41&47.59&52.90&55.99&44.63&49.53&50.77\\%
&&base&49.94&52.74&43.50&32.35&44.63&55.69&60.55&47.35&53.64&54.31\\%
&&novel&68.42&83.86&55.43&53.81&65.38&36.18&28.60&28.30&24.88&29.49\\%
&&$\Delta$&27.01\%&37.11\%&21.53\%&39.89\%&31.74\%&{-}53.91\%&{-}111.71\%&{-}67.33\%&{-}115.65\%&{-}84.16\%\\%
\cline{2%
-%
13}%
&\multirow{4}{*}{3d\_con\_align}&all&60.64&60.41&47.82&38.91&51.95&21.63&29.19&27.08&25.66&25.89\\%
&&base&60.03&59.82&47.02&38.54&51.35&21.80&29.89&27.55&26.07&26.33\\%
&&novel&64.28&63.94&52.66&41.12&55.50&20.59&24.99&24.27&23.22&23.27\\%
&&$\Delta$&6.61\%&6.44\%&10.71\%&6.27\%&7.47\%&{-}5.88\%&{-}19.59\%&{-}13.52\%&{-}12.26\%&{-}13.15\%\\%
\cline{2%
-%
13}%
&\multirow{2}{*}{localnet}&all&71.93&71.47&50.99&43.23&59.40&10.93&21.92&15.91&16.36&16.28\\%
&&base&70.64&68.98&48.68&41.90&57.55&11.42&24.53&16.81&17.05&17.45\\%
&\multirow{2}{*}{(supervised)}&novel&79.66&86.41&64.84&51.21&70.53&7.99&6.25&10.48&12.23&9.24\\%
&&$\Delta$&11.32\%&20.18\%&24.92\%&18.19\%&18.41\%&{-}42.99\%&{-}292.69\%&{-}60.36\%&{-}39.35\%&{-}88.93\%\\%
\hline%
\multirow{12}{*}{3}&\multirow{4}{*}{3d\_finetune}&all&53.92&60.16&47.94&36.24&49.57&52.14&52.74&40.64&49.66&48.80\\%
&&base&51.61&56.12&46.51&33.30&46.88&54.27&57.69&43.23&52.82&52.00\\%
&&novel&67.80&84.38&56.56&53.90&65.66&39.34&23.04&25.12&30.69&29.54\\%
&&$\Delta$&23.88\%&33.50\%&17.77\%&38.23\%&28.60\%&{-}37.96\%&{-}150.45\%&{-}72.11\%&{-}72.12\%&{-}76.02\%\\%
\cline{2%
-%
13}%
&\multirow{4}{*}{3d\_con\_align}&all&61.36&60.93&48.53&38.55&52.34&21.59&28.81&26.66&25.66&25.68\\%
&&base&61.06&60.27&47.99&38.25&51.89&21.64&29.42&26.87&26.06&26.00\\%
&&novel&63.13&64.88&51.80&40.33&55.03&21.33&25.15&25.42&23.28&23.80\\%
&&$\Delta$&3.28\%&7.10\%&7.35\%&5.16\%&5.71\%&{-}1.43\%&{-}16.96\%&{-}5.68\%&{-}11.93\%&{-}9.24\%\\%
\cline{2%
-%
13}%
&\multirow{2}{*}{localnet}&all&78.07&74.86&56.98&47.94&64.46&12.73&16.10&16.97&19.12&16.23\\%
&&base&77.57&73.08&55.73&47.09&63.37&13.50&17.41&17.89&19.98&17.19\\%
&\multirow{2}{*}{(supervised)}&novel&81.09&85.53&64.48&53.06&71.04&8.11&8.23&11.46&13.98&10.45\\%
&&$\Delta$&4.35\%&14.55\%&13.57\%&11.24\%&10.80\%&{-}66.36\%&{-}111.55\%&{-}56.08\%&{-}42.92\%&{-}64.60\%\\%
\hline%
\multirow{12}{*}{4}&\multirow{4}{*}{3d\_finetune}&all&55.53&61.34&48.17&36.85&50.47&50.70&49.19&39.65&47.11&46.66\\%
&&base&52.93&56.80&46.32&33.40&47.36&53.41&54.70&43.12&51.48&50.68\\%
&&novel&68.50&84.07&57.37&54.06&66.00&37.17&21.62&22.29&25.30&26.60\\%
&&$\Delta$&22.73\%&32.44\%&19.25\%&38.21\%&28.24\%&{-}43.66\%&{-}153.05\%&{-}93.42\%&{-}103.48\%&{-}90.54\%\\%
\cline{2%
-%
13}%
&\multirow{4}{*}{3d\_con\_align}&all&61.33&60.63&49.52&38.16&52.41&21.46&29.59&26.14&26.32&25.88\\%
&&base&60.69&59.91&49.01&37.56&51.79&21.64&30.34&26.64&27.06&26.42\\%
&&novel&64.52&64.23&52.08&41.19&55.50&20.57&25.80&23.64&22.63&23.16\\%
&&$\Delta$&5.94\%&6.73\%&5.90\%&8.81\%&6.69\%&{-}5.19\%&{-}17.63\%&{-}12.69\%&{-}19.57\%&{-}14.08\%\\%
\cline{2%
-%
13}%
&\multirow{2}{*}{localnet}&all&77.64&81.89&60.06&51.96&67.89&8.95&9.62&12.77&13.38&11.18\\%
&&base&76.73&81.04&58.84&51.44&67.01&9.50&10.28&13.32&13.75&11.71\\%
&\multirow{2}{*}{(supervised)}&novel&82.14&86.15&66.18&54.57&72.26&6.23&6.33&10.02&11.54&8.53\\%
&&$\Delta$&6.58\%&5.94\%&11.09\%&5.73\%&7.26\%&{-}52.50\%&{-}62.41\%&{-}32.94\%&{-}19.22\%&{-}37.34\%\\%
\hline%
\end{tabular}
    \label{tab:kshot}
\end{table*}

\subsection{Ablation Studies}
\label{sec:ablation_studies}

\textbf{Ablation on different modules}
To assess the effectiveness of different modules in the proposed method, we report the results of the following variations.
\begin{enumerate}
    \item The `3d' network variant - The proposed 3D local prototypical network, detailed in Section~\ref{sec:local_prototypical_network} and Fig.~\ref{fig:3d}, without the support mask conditioning or spatial registration.
    \item The `3d\_con' network variant - The `3d' version with the support mask conditioning, but without the spatial registration, described in Section~\ref{sec:shape_aware} and Fig.~\ref{fig:3d_con}). 
    \item The `3d\_align' network - The `3d' version of the proposed network (detailed in \ref{fig:3d_align}) with the spatial registration mechanism, without the support mask conditioning.
    \item The `3d\_con\_align' network - The ``complete'' version of the proposed network with both the support mask conditioning module and the spatial registration mechanism, as shown in Fig.~\ref{fig:3d_con_align}.
\end{enumerate}

\textbf{Ablation study on the number of shots}
We report model performance when different number of support trios available in each episode. 

\textbf{Ablation on varying training data availability} 
To investigate the dependency of our proposed method on the training set, we report the performance of the proposed model (3d\_con\_align) trained on various availability in training sets. The `half' and `quarter' experiments respectively includes $\frac{1}{2}$ and $\frac{1}{4}$ of the training subset $\mathcal{D}^{train}_u$ for each base institutions $u\in\mathcal{U}_{base}$, in order to test the impact of the training data set size on the few-shot segmentation performance. The same ratio between institutions was maintained in these experiments. Additionally, the `half\_single\_ins' experiment tests the same number of images as the `half' experiment, but all the images are sampled from the same institution (Institution 1).

To assess and compare the intra- and inter-institution generlisation, results are also reported when all the institutions, only the base institutions and only the novel institutions were used as support institutions, denoted as `all', `base' and `novel', respectively.

\section{Results}
The Dice score and 95\% Hausdorff distance achieved by variations of our proposed method as well as the baseline networks are presented for comparison.
Table~\ref{tab:Dicethree} and ~\ref{tab:Dicefour} summarise the network performances, with respect to different folds and the mean of all folds using institution 3 and 4 as the novel institution, respectively. Other institutions were not used as novel institution as they have either too many or too few samples such that the training or test set size will be too small.
When Institution 3 is used as the novel institution (Table~\ref{tab:Dicethree}), the proposed method with both support mask conditioning module and spatial registration mechanism (`3d\_con\_align') outperformed `BiGRU' by 16.34\%, 16.16\% and 17.44\%, and outperformed the `2d' baseline by 11.39\%, 11.79\% and 8.95\% in absolute Dice improvement when support images came from all, base and novel institutions, respectively.
\textcolor{black}{While the `3d\_finetune' baseline achieved higher Dice than the `2d' baseline, its 95\% Hausdorff distance is more than double of the `2d' baseline. This could be related to the commonly appearing false positive predictions as shown in Fig.~\ref{fig:vis}. Dice may still have small yet distant false positive predictions which caused higher Hausdorff distance. We refer the reader to our recent study for a discussion on this particular issue~\citep{yan2022impact}.}
Similar improvements have also been observed when Institution 4 was adopted as the novel institution. $8.64\%$, $8.55\%$ and $9.15\%$ absolute Dice improvements over the `2d' baseline were achieved by the proposed method, as support images came from all, base and novel institutions, respectively (Table~\ref{tab:Dicefour}).
The relative difference between performances when support images come from base and novel institutions, denoted as performance gap $\Delta$, are also summarised in Table~\ref{tab:Dicethree} and Table~\ref{tab:Dicefour}. While achieving better performance overall, the proposed method reported smaller performance gap.

\begin{figure}[h]
    \centering
    \subfigure[]{
    \includegraphics[width=75mm]{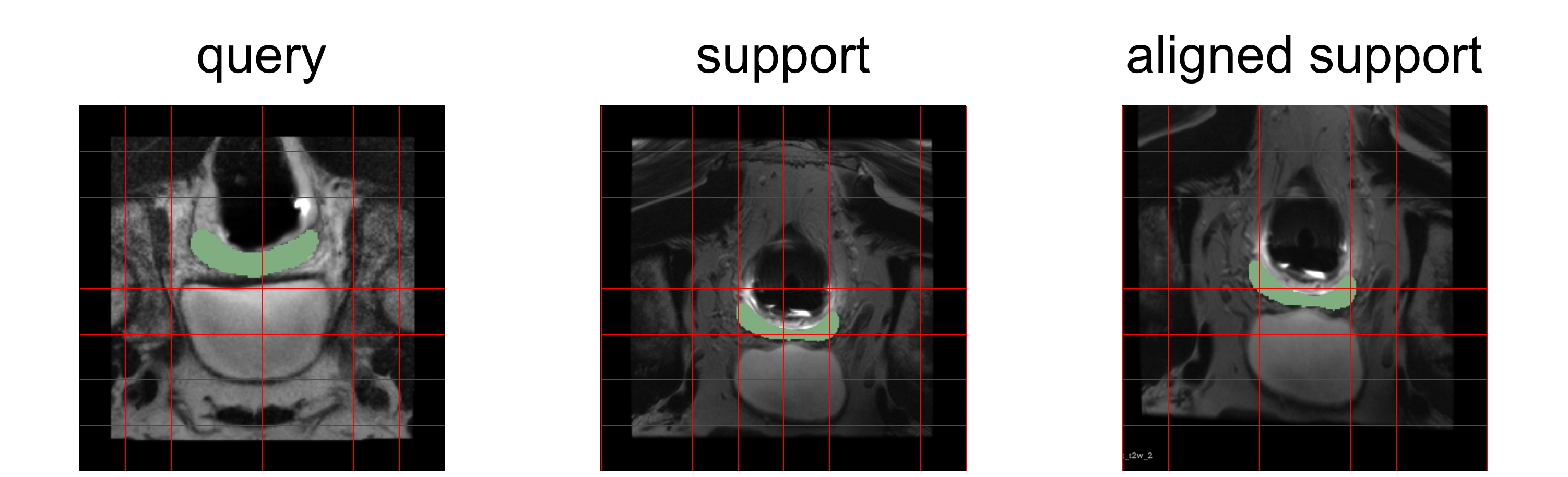}
    }
    \subfigure[]{
    \includegraphics[width=75mm]{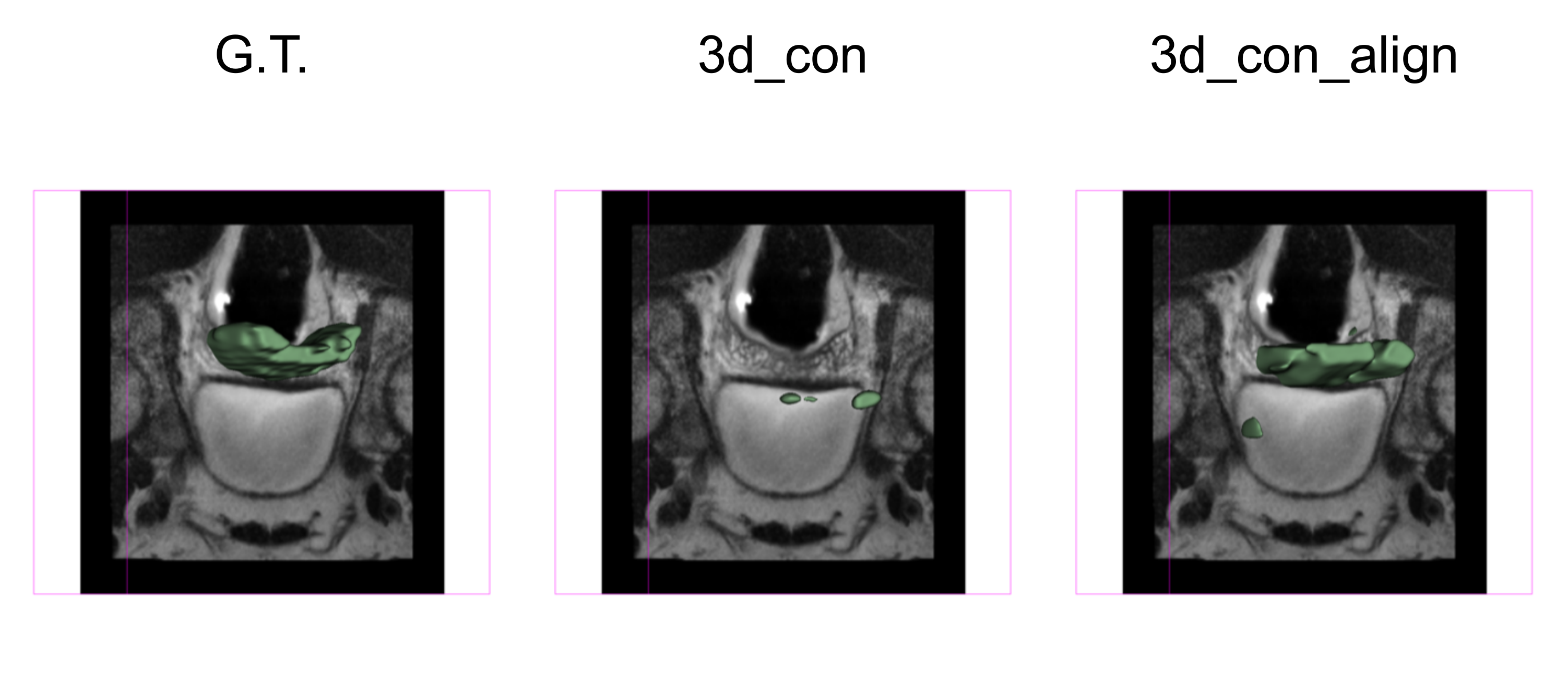}
    }
\caption{\label{fig:align_vis} Qualitative results achieved by `3d\_con' and `3d\_con\_align' on a misaligned query-support pair. (a) visualises the original query and support, as well as the support aligned towards query through the spatial registration mechanism. (b) visualises predictions made by `3d\_con' and `3d\_con\_align' based on the same support-query pair.}
\end{figure}
\begin{figure}[h]
\centering
\includegraphics[width=75mm]{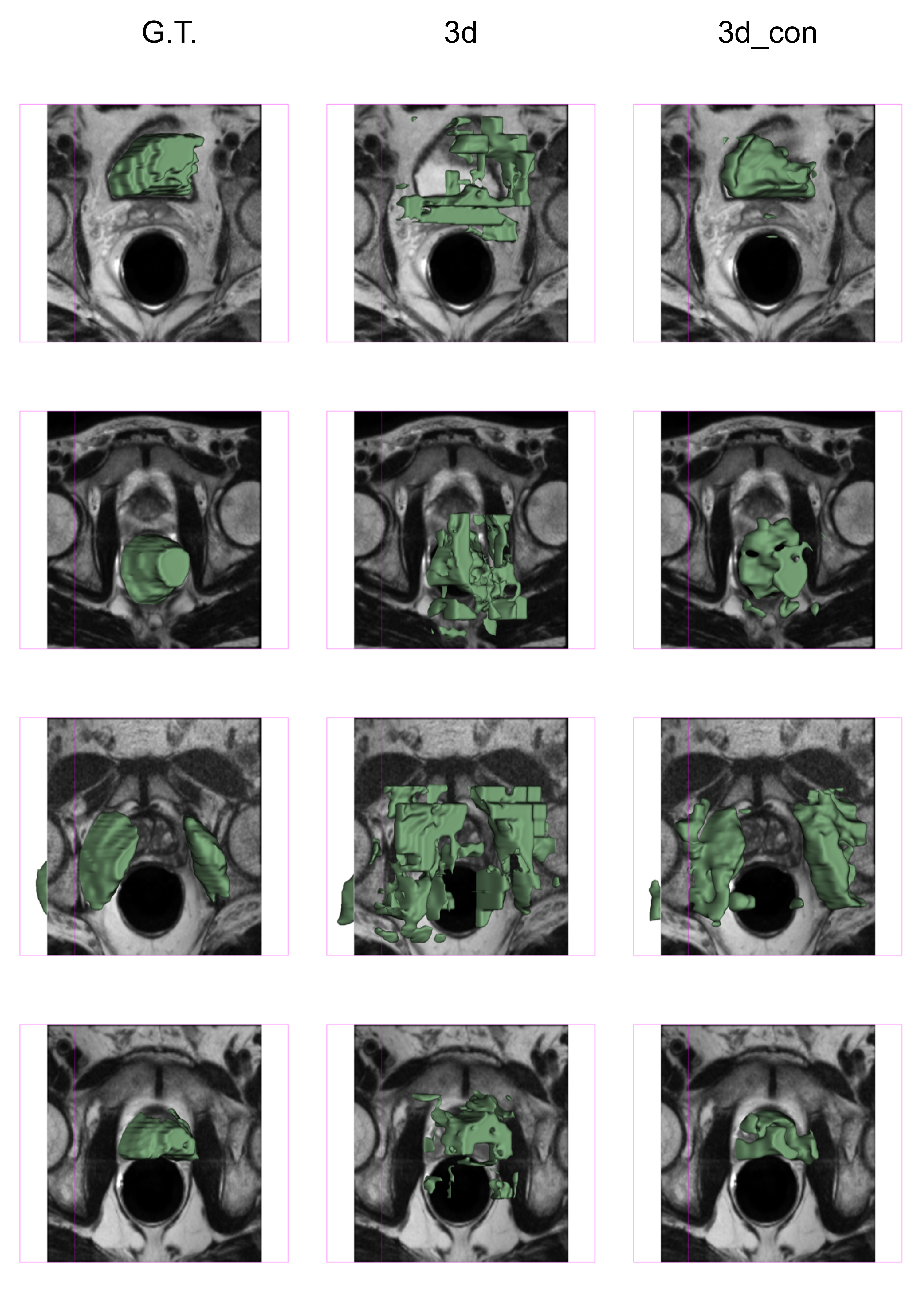}
\caption{\label{fig:con_vis}Qualitative results achieved by `3d' and `3d\_con' on the same support-query pairs. `3d\_con' resulted in more compact segmentation with smoother boundary compare with `3d'}
\end{figure}


\textcolor{black}{Adding support mask conditioning module alone (`3d\_con') led $3.76\%$, $3.25\%$ and $6.76\%$ absolute increase in Dice score comparing to `3d', when support images came from all, base and novel institutions, respectively. Qualitatively, it predicted more ``compact'' segmentation with smoother boundary as shown in Fig.~9. However, it also resulted in a higher $\Delta$ - a greater improvement was achieved when support and query came from the same institution, possibly because of its sensitivity to support-query (mis)alignment.}

This was mitigated by the spatial registration mechanism which not only further improved the Dice score by $6.81\%$ but also reduced $\Delta$ by $9.00\%$. Fig.~10 shows an example where the target structure of the query and support were misaligned, resulting in segmentation failure. Aligning the support towards query, the spatial registration mechanism considerably improved the performance.

\textcolor{black}{Interestingly, when spatial alignment mechanism was available, the support mask conditioning module led to a further improvement - $9.36\%$, $9.35\%$ and $9.41\%$ absolute increase in Dice score from `3d\_align' to `3d\_con\_align' when support images came from all, base and novel institutions, respectively.}

It is also important to report that the proposed 3D network `3d' without the support mask conditioning module and the spatial registration mechanism contained $5.7$ million parameters. It achieved comparable performance to the `2d' baseline with $23.5$ million parameters, resulted in around $75\%$ reduction in number of parameters. For reference, the complete version of our proposed method `3d\_con\_align' had contained 27.3 million parameters ($16\%$ more parameters compared with `2d'), had achieved $28.9\%$ of relative Dice improvement over `2d'.

Furthermore, Table~\ref{tab:cross_2d} and Table~\ref{tab:cross_ours} report the mean Dice score achieved by `2d' and `3d\_con\_align' from different support-query institution combinations, when institution 3 is the novel institution. 
Better performance was often achieved by both methods when support images come from the query institution. A two tailed t-test is performed per query institution between the Dice scores where support institution equals query institution and the maximum Dice scores achieved when support institution differs from the query institution.
Such observation is consistent with the hypothesis that the domain shift is smaller between support-query pairs within the same institution, leading to better performance. 

Table~\ref{tab:BaseIns} reports the performance of the proposed model (`3d\_con\_align') and baseline methods on novel classes, when both query and support come from base institutions while institution 3 is used as the novel institution. We report results when support come from the same institution as query, a different institution from query, and over all base institutions.
The proposed method outperformed all baseline methods in few-shot setting, including two state-of-the-art methods - BiGRU~\citep{kim2021bidirectional} by 17.01\%, 16.88\% and 16.9\% and LSNet~\citep{yu2021location} (which is adopted as our `2d' baseline) by 9.39\%, 13.37\% and 12.88\% absolute Dice, when support come from the same institution as query, different institution from query and all base institutions respectively, proving its efficacy even when no novel institution is involved during evaluation.
\textcolor{black}{Moreover, better performance was achieved when query and support came from the same institution than different institutions, with the performance gap reported as $\Delta$. The proposed method achieved smaller mean $\Delta$ compare to baseline few-shot methods, suggesting its ability to mitigate domain shifts from cross-institution query and support data.}

Table~\ref{tab:trainsize} reports the performance achieved by the proposed method (`3d\_con\_align') as training set varies. The performance dropped as the number of images inside the training set reduced. Notably, model trained on `half\_single\_ins' performed worse than model trained on `quarter', which had only half the size of `half\_single\_ins'. This suggests that the cross-institution few-shot task can be sensitive to the number of institutions available in the training set. To quantify this sensitivity is an interesting future research question.

Table~\ref{tab:kshot} reports the performance of the proposed method `3d\_con\_align' and `3d\_finetune' baseline, using 1 to 4 support examples (denoted as \# shot). Performance improved for both methods, as the number of support examples increased. 
While `3d\_finetune' is more sensitive to the number of support examples, the proposed method still outperformed `3d\_finetune' using 1 to 4 support examples.

\section{Discussion}
While both the spatial registration mechanism and the support mask conditioning were motivated by the observation on multiple structure types found in the multi-institution data set used in this study, they may be promising to be beneficial for wider image types and anatomical regions as similar challenges were found in the few-shot segmentation of other types of non-medical data.

The labels used in this study were annotated by a mixture of clinicians and experienced medical imaging researchers. The estimated time for completing this task was more than one thousand observer-hours, a practically challenging task for most local hospitals if an alternative supervised learning was adopted for adopting or validating a segmentation tool. This further justifies the clinical relevance for the proposed few-shot segmentation approach.

It is noteworthy that the reported segmentation performance results were based on as few as 1 - 4 labelled training examples of the regions of interest, which had not been labelled in the model training stage.

Though challenging, cross-institutional few-shot segmentation could benefit situations when limited number of annotated data are available. Potential applications, although not investigated in this study due to relevant data availability, include specific pathology detection and segmentation with rare instances without previously observed occurrence in training institutions and longitudinal analysis with available within-subject data from individual patients.

\textcolor{black}{Research questions remain for future research include the achievability or conditions to fill the gap to the upper-bound performance from supervised learning. For example, whether other types of data variance, such as scanner, imaging protocol and intensity in addition to the spatial domain studied in this work, need to be considered to improve cross-institution generalisability. Future work should aim to address these challenges for better performance, which we believe is very much plausible. These may yet be crucial in achieving clinically required accuracy for potential adoption in specific applications.}

\section{Conclusion}
This paper described the first 3D prototypical learning algorithm for medical image segmentation, applied on multiple structures on pelvic MR images from different institutes. Substantial validation was based on clinical data from $589$ patients, with full segmentation of eight anatomical classes made available to the scientific community. The demonstrated novelty, efficacy, and clinical applicability of the proposed algorithm suggested an interesting direction for addressing the cost of expert labelling and cross-institute generalisation of current deep learning-based segmentation applications.

\section*{Acknowledgments}
This work was supported by the International Alliance for Cancer Early Detection, a partnership between Cancer Research UK [C28070/A30912; C73666/A31378], Canary Center at Stanford University, the University of Cambridge, OHSU Knight Cancer Institute, University College London and the University of Manchester. This work was also supported by 
the Wellcome/EPSRC Centre for Interventional and Surgical Sciences [203145Z/16/Z], and EPSRC [EP/T029404/1, EP/S021930/1].

\bibliographystyle{model2-names.bst}\biboptions{authoryear}
\bibliography{refs}

\begin{thebibliography}{55}
\expandafter\ifx\csname natexlab\endcsname\relax\def\natexlab#1{#1}\fi
\providecommand{\url}[1]{\texttt{#1}}
\providecommand{\href}[2]{#2}
\providecommand{\path}[1]{#1}
\providecommand{\DOIprefix}{doi:}
\providecommand{\ArXivprefix}{arXiv:}
\providecommand{\URLprefix}{URL: }
\providecommand{\Pubmedprefix}{pmid:}
\providecommand{\doi}[1]{\href{http://dx.doi.org/#1}{\path{#1}}}
\providecommand{\Pubmed}[1]{\href{pmid:#1}{\path{#1}}}
\providecommand{\bibinfo}[2]{#2}
\ifx\xfnm\relax \def\xfnm[#1]{\unskip,\space#1}\fi
\bibitem[{Abdel-Basset et~al.(2021)Abdel-Basset, Chang, Hawash, Chakrabortty
  and Ryan}]{abdel2021fss}
\bibinfo{author}{Abdel-Basset, M.}, \bibinfo{author}{Chang, V.},
  \bibinfo{author}{Hawash, H.}, \bibinfo{author}{Chakrabortty, R.K.},
  \bibinfo{author}{Ryan, M.}, \bibinfo{year}{2021}.
\newblock \bibinfo{title}{Fss-2019-ncov: A deep learning architecture for
  semi-supervised few-shot segmentation of covid-19 infection}.
\newblock \bibinfo{journal}{Knowledge-Based Systems} \bibinfo{volume}{212},
  \bibinfo{pages}{106647}.
\bibitem[{Bloch et~al.(2015)Bloch, Jain and Jaffe}]{ProstateDx}
\bibinfo{author}{Bloch, B.N.}, \bibinfo{author}{Jain, A.},
  \bibinfo{author}{Jaffe, C.C.}, \bibinfo{year}{2015}.
\newblock \bibinfo{title}{Data from prostate-diagnosis}.
\newblock \bibinfo{journal}{The Cancer Imaging Archive}
  \DOIprefix\doi{10.7937/K9/TCIA.2015.FOQEUJVT}.
\bibitem[{Cardoso et~al.(2022)Cardoso, Li, Brown, Ma, Kerfoot, Wang, Murrey,
  Myronenko, Zhao, Yang et~al.}]{cardoso2022monai}
\bibinfo{author}{Cardoso, M.J.}, \bibinfo{author}{Li, W.},
  \bibinfo{author}{Brown, R.}, \bibinfo{author}{Ma, N.},
  \bibinfo{author}{Kerfoot, E.}, \bibinfo{author}{Wang, Y.},
  \bibinfo{author}{Murrey, B.}, \bibinfo{author}{Myronenko, A.},
  \bibinfo{author}{Zhao, C.}, \bibinfo{author}{Yang, D.}, et~al.,
  \bibinfo{year}{2022}.
\newblock \bibinfo{title}{Monai: An open-source framework for deep learning in
  healthcare}.
\newblock \bibinfo{journal}{arXiv preprint arXiv:2211.02701} .
\bibitem[{Choyke et~al.(2016)Choyke, Turkbey, Pinto, Merino and
  Wood}]{ProstateMRI}
\bibinfo{author}{Choyke, P.}, \bibinfo{author}{Turkbey, B.},
  \bibinfo{author}{Pinto, P.}, \bibinfo{author}{Merino, M.},
  \bibinfo{author}{Wood, B.}, \bibinfo{year}{2016}.
\newblock \bibinfo{title}{Data from prostate-mri}.
\newblock \bibinfo{journal}{The Cancer Imaging Archive}
  \DOIprefix\doi{10.7937/K9/TCIA.2016.6046GUDv}.
\bibitem[{Cui et~al.(2020)Cui, Wei, Ma, Gu and Zheng}]{cui2020unified}
\bibinfo{author}{Cui, H.}, \bibinfo{author}{Wei, D.}, \bibinfo{author}{Ma, K.},
  \bibinfo{author}{Gu, S.}, \bibinfo{author}{Zheng, Y.}, \bibinfo{year}{2020}.
\newblock \bibinfo{title}{A unified framework for generalized low-shot medical
  image segmentation with scarce data}.
\newblock \bibinfo{journal}{IEEE Transactions on Medical Imaging}
  \bibinfo{volume}{40}, \bibinfo{pages}{2656--2671}.
\bibitem[{Dickinson et~al.(2013)Dickinson, Ahmed, Kirkham, Allen, Freeman,
  Barber, Hindley, Leslie, Ogden, Persad et~al.}]{dickinson2013multi}
\bibinfo{author}{Dickinson, L.}, \bibinfo{author}{Ahmed, H.U.},
  \bibinfo{author}{Kirkham, A.}, \bibinfo{author}{Allen, C.},
  \bibinfo{author}{Freeman, A.}, \bibinfo{author}{Barber, J.},
  \bibinfo{author}{Hindley, R.G.}, \bibinfo{author}{Leslie, T.},
  \bibinfo{author}{Ogden, C.}, \bibinfo{author}{Persad, R.}, et~al.,
  \bibinfo{year}{2013}.
\newblock \bibinfo{title}{A multi-centre prospective development study
  evaluating focal therapy using high intensity focused ultrasound for
  localised prostate cancer: the index study}.
\newblock \bibinfo{journal}{Contemporary clinical trials} \bibinfo{volume}{36},
  \bibinfo{pages}{68--80}.
\bibitem[{Dong and Xing(2018)}]{dong2018few}
\bibinfo{author}{Dong, N.}, \bibinfo{author}{Xing, E.P.}, \bibinfo{year}{2018}.
\newblock \bibinfo{title}{Few-shot semantic segmentation with prototype
  learning.}, in: \bibinfo{booktitle}{BMVC}.
\bibitem[{Feyjie et~al.(2020)Feyjie, Azad, Pedersoli, Kauffman, Ayed and
  Dolz}]{feyjie2020semi}
\bibinfo{author}{Feyjie, A.R.}, \bibinfo{author}{Azad, R.},
  \bibinfo{author}{Pedersoli, M.}, \bibinfo{author}{Kauffman, C.},
  \bibinfo{author}{Ayed, I.B.}, \bibinfo{author}{Dolz, J.},
  \bibinfo{year}{2020}.
\newblock \bibinfo{title}{Semi-supervised few-shot learning for medical image
  segmentation}.
\newblock \bibinfo{journal}{arXiv preprint arXiv:2003.08462} .
\bibitem[{Fiorino et~al.(1998)Fiorino, Reni, Bolognesi, Cattaneo and
  Calandrino}]{fiorino1998intra}
\bibinfo{author}{Fiorino, C.}, \bibinfo{author}{Reni, M.},
  \bibinfo{author}{Bolognesi, A.}, \bibinfo{author}{Cattaneo, G.M.},
  \bibinfo{author}{Calandrino, R.}, \bibinfo{year}{1998}.
\newblock \bibinfo{title}{Intra-and inter-observer variability in contouring
  prostate and seminal vesicles: implications for conformal treatment
  planning}.
\newblock \bibinfo{journal}{Radiotherapy and oncology} \bibinfo{volume}{47},
  \bibinfo{pages}{285--292}.
\bibitem[{Fu et~al.(2019)Fu, Robu, Koo, Schneider, Laarhoven, Stoyanov,
  Davidson, Clarkson and Hu}]{fu2019more}
\bibinfo{author}{Fu, Y.}, \bibinfo{author}{Robu, M.R.}, \bibinfo{author}{Koo,
  B.}, \bibinfo{author}{Schneider, C.}, \bibinfo{author}{Laarhoven, S.v.},
  \bibinfo{author}{Stoyanov, D.}, \bibinfo{author}{Davidson, B.},
  \bibinfo{author}{Clarkson, M.J.}, \bibinfo{author}{Hu, Y.},
  \bibinfo{year}{2019}.
\newblock \bibinfo{title}{More unlabelled data or label more data? a study on
  semi-supervised laparoscopic image segmentation}, in:
  \bibinfo{booktitle}{Domain Adaptation and Representation Transfer and Medical
  Image Learning with Less Labels and Imperfect Data}.
  \bibinfo{publisher}{Springer}, pp. \bibinfo{pages}{173--180}.
\bibitem[{Gibson et~al.(2018)Gibson, Hu, Ghavami et~al.}]{gibson2018inter}
\bibinfo{author}{Gibson, E.}, \bibinfo{author}{Hu, Y.},
  \bibinfo{author}{Ghavami, N.}, et~al., \bibinfo{year}{2018}.
\newblock \bibinfo{title}{Inter-site variability in prostate segmentation
  accuracy using deep learning}, in: \bibinfo{booktitle}{International
  Conference on Medical Image Computing and Computer-Assisted Intervention},
  \bibinfo{organization}{Springer}. pp. \bibinfo{pages}{506--514}.
\bibitem[{Guo et~al.(2021)Guo, Xu, Feng, Xiong, Gao and Zhang}]{guo2021multi}
\bibinfo{author}{Guo, S.}, \bibinfo{author}{Xu, L.}, \bibinfo{author}{Feng,
  C.}, \bibinfo{author}{Xiong, H.}, \bibinfo{author}{Gao, Z.},
  \bibinfo{author}{Zhang, H.}, \bibinfo{year}{2021}.
\newblock \bibinfo{title}{Multi-level semantic adaptation for few-shot
  segmentation on cardiac image sequences}.
\newblock \bibinfo{journal}{Medical Image Analysis} \bibinfo{volume}{73},
  \bibinfo{pages}{102170}.
\bibitem[{Hamid et~al.(2019)Hamid, Donaldson, Hu et~al.}]{hamid2019smarttarget}
\bibinfo{author}{Hamid, S.}, \bibinfo{author}{Donaldson, I.A.},
  \bibinfo{author}{Hu, Y.}, et~al., \bibinfo{year}{2019}.
\newblock \bibinfo{title}{The smarttarget biopsy trial: a prospective,
  within-person randomised, blinded trial comparing the accuracy of
  visual-registration and magnetic resonance imaging/ultrasound image-fusion
  targeted biopsies for prostate cancer risk stratification}.
\newblock \bibinfo{journal}{European urology} \bibinfo{volume}{75},
  \bibinfo{pages}{733--740}.
\bibitem[{Han and Fischl(2007)}]{han2007atlas}
\bibinfo{author}{Han, X.}, \bibinfo{author}{Fischl, B.}, \bibinfo{year}{2007}.
\newblock \bibinfo{title}{Atlas renormalization for improved brain mr image
  segmentation across scanner platforms}.
\newblock \bibinfo{journal}{IEEE transactions on medical imaging}
  \bibinfo{volume}{26}, \bibinfo{pages}{479--486}.
\bibitem[{He et~al.(2020)He, Li, Yang, Kong, Chen, Shu, Coatrieux, Dillenseger
  and Li}]{he2020deep}
\bibinfo{author}{He, Y.}, \bibinfo{author}{Li, T.}, \bibinfo{author}{Yang, G.},
  \bibinfo{author}{Kong, Y.}, \bibinfo{author}{Chen, Y.}, \bibinfo{author}{Shu,
  H.}, \bibinfo{author}{Coatrieux, J.L.}, \bibinfo{author}{Dillenseger, J.L.},
  \bibinfo{author}{Li, S.}, \bibinfo{year}{2020}.
\newblock \bibinfo{title}{Deep complementary joint model for complex scene
  registration and few-shot segmentation on medical images}, in:
  \bibinfo{booktitle}{European Conference on Computer Vision},
  \bibinfo{organization}{Springer}. pp. \bibinfo{pages}{770--786}.
\bibitem[{Henschel et~al.(2020)Henschel, Conjeti, Estrada, Diers, Fischl and
  Reuter}]{henschel2020fastsurfer}
\bibinfo{author}{Henschel, L.}, \bibinfo{author}{Conjeti, S.},
  \bibinfo{author}{Estrada, S.}, \bibinfo{author}{Diers, K.},
  \bibinfo{author}{Fischl, B.}, \bibinfo{author}{Reuter, M.},
  \bibinfo{year}{2020}.
\newblock \bibinfo{title}{Fastsurfer-a fast and accurate deep learning based
  neuroimaging pipeline}.
\newblock \bibinfo{journal}{NeuroImage} \bibinfo{volume}{219},
  \bibinfo{pages}{117012}.
\bibitem[{Hosseini-Asl et~al.(2016)Hosseini-Asl, Keynton and
  El-Baz}]{hosseini2016alzheimer}
\bibinfo{author}{Hosseini-Asl, E.}, \bibinfo{author}{Keynton, R.},
  \bibinfo{author}{El-Baz, A.}, \bibinfo{year}{2016}.
\newblock \bibinfo{title}{Alzheimer's disease diagnostics by adaptation of 3d
  convolutional network}, in: \bibinfo{booktitle}{2016 IEEE international
  conference on image processing (ICIP)}, \bibinfo{organization}{IEEE}. pp.
  \bibinfo{pages}{126--130}.
\bibitem[{Howe and Matsuoka(1999)}]{howe1999robotics}
\bibinfo{author}{Howe, R.D.}, \bibinfo{author}{Matsuoka, Y.},
  \bibinfo{year}{1999}.
\newblock \bibinfo{title}{Robotics for surgery}.
\newblock \bibinfo{journal}{Annual review of biomedical engineering}
  \bibinfo{volume}{1}, \bibinfo{pages}{211--240}.
\bibitem[{Hu et~al.(2018a)Hu, Modat, Gibson, Ghavami, Bonmati, Moore, Emberton,
  Noble, Barratt and Vercauteren}]{hu2018label}
\bibinfo{author}{Hu, Y.}, \bibinfo{author}{Modat, M.}, \bibinfo{author}{Gibson,
  E.}, \bibinfo{author}{Ghavami, N.}, \bibinfo{author}{Bonmati, E.},
  \bibinfo{author}{Moore, C.M.}, \bibinfo{author}{Emberton, M.},
  \bibinfo{author}{Noble, J.A.}, \bibinfo{author}{Barratt, D.C.},
  \bibinfo{author}{Vercauteren, T.}, \bibinfo{year}{2018}a.
\newblock \bibinfo{title}{Label-driven weakly-supervised learning for
  multimodal deformable image registration}, in: \bibinfo{booktitle}{2018 IEEE
  15th International Symposium on Biomedical Imaging (ISBI 2018)},
  \bibinfo{organization}{IEEE}. pp. \bibinfo{pages}{1070--1074}.
\bibitem[{Hu et~al.(2018b)Hu, Modat, Gibson, Li, Ghavami, Bonmati, Wang,
  Bandula, Moore, Emberton et~al.}]{hu2018weakly}
\bibinfo{author}{Hu, Y.}, \bibinfo{author}{Modat, M.}, \bibinfo{author}{Gibson,
  E.}, \bibinfo{author}{Li, W.}, \bibinfo{author}{Ghavami, N.},
  \bibinfo{author}{Bonmati, E.}, \bibinfo{author}{Wang, G.},
  \bibinfo{author}{Bandula, S.}, \bibinfo{author}{Moore, C.M.},
  \bibinfo{author}{Emberton, M.}, et~al., \bibinfo{year}{2018}b.
\newblock \bibinfo{title}{Weakly-supervised convolutional neural networks for
  multimodal image registration}.
\newblock \bibinfo{journal}{Medical image analysis} \bibinfo{volume}{49},
  \bibinfo{pages}{1--13}.
\bibitem[{Hutchinson and Raff(2000)}]{hutchinson2000structural}
\bibinfo{author}{Hutchinson, M.}, \bibinfo{author}{Raff, U.},
  \bibinfo{year}{2000}.
\newblock \bibinfo{title}{Structural changes of the substantia nigra in
  parkinson's disease as revealed by mr imaging}.
\newblock \bibinfo{journal}{American journal of neuroradiology}
  \bibinfo{volume}{21}, \bibinfo{pages}{697--701}.
\bibitem[{Kim et~al.(2021)Kim, An, Chikontwe and Park}]{kim2021bidirectional}
\bibinfo{author}{Kim, S.}, \bibinfo{author}{An, S.},
  \bibinfo{author}{Chikontwe, P.}, \bibinfo{author}{Park, S.H.},
  \bibinfo{year}{2021}.
\newblock \bibinfo{title}{Bidirectional rnn-based few shot learning for 3d
  medical image segmentation}, in: \bibinfo{booktitle}{Proceedings of the AAAI
  Conference on Artificial Intelligence}, pp. \bibinfo{pages}{1808--1816}.
\bibitem[{Li et~al.(2020)Li, Wang, Li, Liu and Zhang}]{li2020predicting}
\bibinfo{author}{Li, A.}, \bibinfo{author}{Wang, S.}, \bibinfo{author}{Li, W.},
  \bibinfo{author}{Liu, S.}, \bibinfo{author}{Zhang, S.}, \bibinfo{year}{2020}.
\newblock \bibinfo{title}{Predicting human mobility with federated learning},
  in: \bibinfo{booktitle}{Proceedings of the 28th International Conference on
  Advances in Geographic Information Systems}, pp. \bibinfo{pages}{441--444}.
\bibitem[{Li et~al.(2021a)Li, Jiang, Zhang, Kamp and Dou}]{li2021fedbn}
\bibinfo{author}{Li, X.}, \bibinfo{author}{Jiang, M.}, \bibinfo{author}{Zhang,
  X.}, \bibinfo{author}{Kamp, M.}, \bibinfo{author}{Dou, Q.},
  \bibinfo{year}{2021}a.
\newblock \bibinfo{title}{Fedbn: Federated learning on non-iid features via
  local batch normalization}.
\newblock \bibinfo{journal}{arXiv preprint arXiv:2102.07623} .
\bibitem[{Li et~al.(2021b)Li, Data, Fu, Hu and Prisacariu}]{li2021few}
\bibinfo{author}{Li, Y.}, \bibinfo{author}{Data, G.W.P.}, \bibinfo{author}{Fu,
  Y.}, \bibinfo{author}{Hu, Y.}, \bibinfo{author}{Prisacariu, V.A.},
  \bibinfo{year}{2021}b.
\newblock \bibinfo{title}{Few-shot semantic segmentation with self-supervision
  from pseudo-classes}.
\newblock \bibinfo{journal}{arXiv preprint arXiv:2110.11742} .
\bibitem[{Li et~al.(2022)Li, Fu, Yang, Min, Yan, Huisman, Barratt, Prisacariu
  and Hu}]{li2022few}
\bibinfo{author}{Li, Y.}, \bibinfo{author}{Fu, Y.}, \bibinfo{author}{Yang, Q.},
  \bibinfo{author}{Min, Z.}, \bibinfo{author}{Yan, W.},
  \bibinfo{author}{Huisman, H.}, \bibinfo{author}{Barratt, D.},
  \bibinfo{author}{Prisacariu, V.A.}, \bibinfo{author}{Hu, Y.},
  \bibinfo{year}{2022}.
\newblock \bibinfo{title}{Few-shot image segmentation for cross-institution
  male pelvic organs using registration-assisted prototypical learning}, in:
  \bibinfo{booktitle}{2022 IEEE 19th International Symposium on Biomedical
  Imaging (ISBI)}, \bibinfo{organization}{IEEE}. pp. \bibinfo{pages}{1--5}.
\bibitem[{Litjens et~al.(2015)Litjens, Futterer and Huisman}]{Prostate-3T}
\bibinfo{author}{Litjens, G.}, \bibinfo{author}{Futterer, J.},
  \bibinfo{author}{Huisman, H.}, \bibinfo{year}{2015}.
\newblock \bibinfo{title}{Data from prostate-3t}.
\newblock \bibinfo{journal}{The Cancer Imaging Archive}
  \DOIprefix\doi{10.7937/K9/TCIA.2015.QJTV5IL5}.
\bibitem[{Litjens et~al.(2014)Litjens, Toth, van~de Ven
  et~al.}]{litjens2014evaluation}
\bibinfo{author}{Litjens, G.}, \bibinfo{author}{Toth, R.},
  \bibinfo{author}{van~de Ven, W.}, et~al., \bibinfo{year}{2014}.
\newblock \bibinfo{title}{Evaluation of prostate segmentation algorithms for
  mri: the promise12 challenge}.
\newblock \bibinfo{journal}{Medical image analysis} \bibinfo{volume}{18},
  \bibinfo{pages}{359--373}.
\bibitem[{Liu et~al.(2020)Liu, Zhang, Zhang and He}]{liu2020part}
\bibinfo{author}{Liu, Y.}, \bibinfo{author}{Zhang, X.}, \bibinfo{author}{Zhang,
  S.}, \bibinfo{author}{He, X.}, \bibinfo{year}{2020}.
\newblock \bibinfo{title}{Part-aware prototype network for few-shot semantic
  segmentation}, in: \bibinfo{booktitle}{European Conference on Computer
  Vision}, \bibinfo{organization}{Springer}. pp. \bibinfo{pages}{142--158}.
\bibitem[{Meyer et~al.(2021)Meyer, Mehrtash, Rak, Bashkanov, Langbein, Ziaei,
  Kibel, Tempany, Hansen and Tokuda}]{meyer2021domain}
\bibinfo{author}{Meyer, A.}, \bibinfo{author}{Mehrtash, A.},
  \bibinfo{author}{Rak, M.}, \bibinfo{author}{Bashkanov, O.},
  \bibinfo{author}{Langbein, B.}, \bibinfo{author}{Ziaei, A.},
  \bibinfo{author}{Kibel, A.S.}, \bibinfo{author}{Tempany, C.M.},
  \bibinfo{author}{Hansen, C.}, \bibinfo{author}{Tokuda, J.},
  \bibinfo{year}{2021}.
\newblock \bibinfo{title}{Domain adaptation for segmentation of critical
  structures for prostate cancer therapy}.
\newblock \bibinfo{journal}{Scientific reports} \bibinfo{volume}{11},
  \bibinfo{pages}{1--14}.
\bibitem[{Mondal et~al.(2018)Mondal, Dolz and Desrosiers}]{mondal2018few}
\bibinfo{author}{Mondal, A.K.}, \bibinfo{author}{Dolz, J.},
  \bibinfo{author}{Desrosiers, C.}, \bibinfo{year}{2018}.
\newblock \bibinfo{title}{Few-shot 3d multi-modal medical image segmentation
  using generative adversarial learning}.
\newblock \bibinfo{journal}{arXiv preprint arXiv:1810.12241} .
\bibitem[{Ouyang et~al.(2020)Ouyang, Biffi, Chen et~al.}]{ouyang2020self}
\bibinfo{author}{Ouyang, C.}, \bibinfo{author}{Biffi, C.},
  \bibinfo{author}{Chen, C.}, et~al., \bibinfo{year}{2020}.
\newblock \bibinfo{title}{Self-supervision with superpixels: Training few-shot
  medical image segmentation without annotation}, in:
  \bibinfo{booktitle}{European Conference on Computer Vision},
  \bibinfo{organization}{Springer}. pp. \bibinfo{pages}{762--780}.
\bibitem[{Perone et~al.(2019)Perone, Ballester, Barros and
  Cohen-Adad}]{perone2019unsupervised}
\bibinfo{author}{Perone, C.S.}, \bibinfo{author}{Ballester, P.},
  \bibinfo{author}{Barros, R.C.}, \bibinfo{author}{Cohen-Adad, J.},
  \bibinfo{year}{2019}.
\newblock \bibinfo{title}{Unsupervised domain adaptation for medical imaging
  segmentation with self-ensembling}.
\newblock \bibinfo{journal}{NeuroImage} \bibinfo{volume}{194},
  \bibinfo{pages}{1--11}.
\bibitem[{Petrella et~al.(2003)Petrella, Coleman and
  Doraiswamy}]{petrella2003neuroimaging}
\bibinfo{author}{Petrella, J.R.}, \bibinfo{author}{Coleman, R.E.},
  \bibinfo{author}{Doraiswamy, P.M.}, \bibinfo{year}{2003}.
\newblock \bibinfo{title}{Neuroimaging and early diagnosis of alzheimer
  disease: a look to the future}.
\newblock \bibinfo{journal}{Radiology} \bibinfo{volume}{226},
  \bibinfo{pages}{315--336}.
\bibitem[{Ren et~al.(2018)Ren, Hacihaliloglu, Singer, Foran and
  Qi}]{ren2018adversarial}
\bibinfo{author}{Ren, J.}, \bibinfo{author}{Hacihaliloglu, I.},
  \bibinfo{author}{Singer, E.A.}, \bibinfo{author}{Foran, D.J.},
  \bibinfo{author}{Qi, X.}, \bibinfo{year}{2018}.
\newblock \bibinfo{title}{Adversarial domain adaptation for classification of
  prostate histopathology whole-slide images}, in:
  \bibinfo{booktitle}{International conference on medical image computing and
  computer-assisted intervention}, \bibinfo{organization}{Springer}. pp.
  \bibinfo{pages}{201--209}.
\bibitem[{Rieke et~al.(2020)Rieke, Hancox, Li, Milletari, Roth, Albarqouni,
  Bakas, Galtier, Landman, Maier-Hein et~al.}]{rieke2020future}
\bibinfo{author}{Rieke, N.}, \bibinfo{author}{Hancox, J.}, \bibinfo{author}{Li,
  W.}, \bibinfo{author}{Milletari, F.}, \bibinfo{author}{Roth, H.R.},
  \bibinfo{author}{Albarqouni, S.}, \bibinfo{author}{Bakas, S.},
  \bibinfo{author}{Galtier, M.N.}, \bibinfo{author}{Landman, B.A.},
  \bibinfo{author}{Maier-Hein, K.}, et~al., \bibinfo{year}{2020}.
\newblock \bibinfo{title}{The future of digital health with federated
  learning}.
\newblock \bibinfo{journal}{NPJ digital medicine} \bibinfo{volume}{3},
  \bibinfo{pages}{1--7}.
\bibitem[{De~la Rosette et~al.(2010)De~la Rosette, Ahmed, Barentsz, Johansen,
  Brausi, Emberton, Frauscher, Greene, Harisinghani, Haustermans
  et~al.}]{de2010focal}
\bibinfo{author}{De~la Rosette, J.}, \bibinfo{author}{Ahmed, H.},
  \bibinfo{author}{Barentsz, J.}, \bibinfo{author}{Johansen, T.B.},
  \bibinfo{author}{Brausi, M.}, \bibinfo{author}{Emberton, M.},
  \bibinfo{author}{Frauscher, F.}, \bibinfo{author}{Greene, D.},
  \bibinfo{author}{Harisinghani, M.}, \bibinfo{author}{Haustermans, K.},
  et~al., \bibinfo{year}{2010}.
\newblock \bibinfo{title}{Focal therapy in prostate cancer—report from a
  consensus panel}.
\newblock \bibinfo{journal}{Journal of Endourology} \bibinfo{volume}{24},
  \bibinfo{pages}{775--780}.
\bibitem[{Roy et~al.(2020)Roy, Siddiqui, P{\"o}lsterl et~al.}]{roy2020squeeze}
\bibinfo{author}{Roy, A.G.}, \bibinfo{author}{Siddiqui, S.},
  \bibinfo{author}{P{\"o}lsterl, S.}, et~al., \bibinfo{year}{2020}.
\newblock \bibinfo{title}{‘squeeze \& excite’guided few-shot segmentation
  of volumetric images}.
\newblock \bibinfo{journal}{Medical image analysis} \bibinfo{volume}{59},
  \bibinfo{pages}{101587}.
\bibitem[{Shaban et~al.(2017)Shaban, Bansal, Liu et~al.}]{shaban2017one}
\bibinfo{author}{Shaban, A.}, \bibinfo{author}{Bansal, S.},
  \bibinfo{author}{Liu, Z.}, et~al., \bibinfo{year}{2017}.
\newblock \bibinfo{title}{One-shot learning for semantic segmentation}.
\newblock \bibinfo{journal}{arXiv preprint arXiv:1709.03410} .
\bibitem[{Simmons et~al.(2014)Simmons, Ahmed, Moore, Punwani, Freeman, Hu,
  Barratt, Charman, Van~der Meulen and Emberton}]{simmons2014picture}
\bibinfo{author}{Simmons, L.A.}, \bibinfo{author}{Ahmed, H.U.},
  \bibinfo{author}{Moore, C.M.}, \bibinfo{author}{Punwani, S.},
  \bibinfo{author}{Freeman, A.}, \bibinfo{author}{Hu, Y.},
  \bibinfo{author}{Barratt, D.}, \bibinfo{author}{Charman, S.C.},
  \bibinfo{author}{Van~der Meulen, J.}, \bibinfo{author}{Emberton, M.},
  \bibinfo{year}{2014}.
\newblock \bibinfo{title}{The picture study—prostate imaging
  (multi-parametric mri and prostate histoscanning™) compared to
  transperineal ultrasound guided biopsy for significant prostate cancer risk
  evaluation}.
\newblock \bibinfo{journal}{Contemporary clinical trials} \bibinfo{volume}{37},
  \bibinfo{pages}{69--83}.
\bibitem[{Snell et~al.(2017)Snell, Swersky and Zemel}]{snell2017prototypical}
\bibinfo{author}{Snell, J.}, \bibinfo{author}{Swersky, K.},
  \bibinfo{author}{Zemel, R.}, \bibinfo{year}{2017}.
\newblock \bibinfo{title}{Prototypical networks for few-shot learning}.
\newblock \bibinfo{journal}{Advances in neural information processing systems}
  \bibinfo{volume}{30}.
\bibitem[{Sudre et~al.(2017)Sudre, Li, Vercauteren, Ourselin and
  Jorge~Cardoso}]{sudre2017generalised}
\bibinfo{author}{Sudre, C.H.}, \bibinfo{author}{Li, W.},
  \bibinfo{author}{Vercauteren, T.}, \bibinfo{author}{Ourselin, S.},
  \bibinfo{author}{Jorge~Cardoso, M.}, \bibinfo{year}{2017}.
\newblock \bibinfo{title}{Generalised dice overlap as a deep learning loss
  function for highly unbalanced segmentations}, in: \bibinfo{booktitle}{Deep
  learning in medical image analysis and multimodal learning for clinical
  decision support}. \bibinfo{publisher}{Springer}, pp.
  \bibinfo{pages}{240--248}.
\bibitem[{Sun et~al.(2022)Sun, Li, Ding, Huang, Chen, Wang, Yu and
  Paisley}]{sun2022few}
\bibinfo{author}{Sun, L.}, \bibinfo{author}{Li, C.}, \bibinfo{author}{Ding,
  X.}, \bibinfo{author}{Huang, Y.}, \bibinfo{author}{Chen, Z.},
  \bibinfo{author}{Wang, G.}, \bibinfo{author}{Yu, Y.},
  \bibinfo{author}{Paisley, J.}, \bibinfo{year}{2022}.
\newblock \bibinfo{title}{Few-shot medical image segmentation using a global
  correlation network with discriminative embedding}.
\newblock \bibinfo{journal}{Computers in biology and medicine}
  \bibinfo{volume}{140}, \bibinfo{pages}{105067}.
\bibitem[{Sung et~al.(2018)Sung, Yang, Zhang, Xiang, Torr and
  Hospedales}]{sung2018learning}
\bibinfo{author}{Sung, F.}, \bibinfo{author}{Yang, Y.}, \bibinfo{author}{Zhang,
  L.}, \bibinfo{author}{Xiang, T.}, \bibinfo{author}{Torr, P.H.},
  \bibinfo{author}{Hospedales, T.M.}, \bibinfo{year}{2018}.
\newblock \bibinfo{title}{Learning to compare: Relation network for few-shot
  learning}, in: \bibinfo{booktitle}{Proceedings of the IEEE conference on
  computer vision and pattern recognition}, pp. \bibinfo{pages}{1199--1208}.
\bibitem[{Tang et~al.(2021)Tang, Liu, Sun, Yan and Xie}]{tang2021recurrent}
\bibinfo{author}{Tang, H.}, \bibinfo{author}{Liu, X.}, \bibinfo{author}{Sun,
  S.}, \bibinfo{author}{Yan, X.}, \bibinfo{author}{Xie, X.},
  \bibinfo{year}{2021}.
\newblock \bibinfo{title}{Recurrent mask refinement for few-shot medical image
  segmentation}, in: \bibinfo{booktitle}{Proceedings of the IEEE/CVF
  International Conference on Computer Vision}, pp.
  \bibinfo{pages}{3918--3928}.
\bibitem[{Tian et~al.(2020)Tian, Zhao, Shu, Yang, Li and Jia}]{tian2020prior}
\bibinfo{author}{Tian, Z.}, \bibinfo{author}{Zhao, H.}, \bibinfo{author}{Shu,
  M.}, \bibinfo{author}{Yang, Z.}, \bibinfo{author}{Li, R.},
  \bibinfo{author}{Jia, J.}, \bibinfo{year}{2020}.
\newblock \bibinfo{title}{Prior guided feature enrichment network for few-shot
  segmentation}.
\newblock \bibinfo{journal}{IEEE transactions on pattern analysis and machine
  intelligence} .
\bibitem[{Tomar et~al.(2021)Tomar, Lortkipanidze, Vray, Bozorgtabar and
  Thiran}]{tomar2021self}
\bibinfo{author}{Tomar, D.}, \bibinfo{author}{Lortkipanidze, M.},
  \bibinfo{author}{Vray, G.}, \bibinfo{author}{Bozorgtabar, B.},
  \bibinfo{author}{Thiran, J.P.}, \bibinfo{year}{2021}.
\newblock \bibinfo{title}{Self-attentive spatial adaptive normalization for
  cross-modality domain adaptation}.
\newblock \bibinfo{journal}{IEEE Transactions on Medical Imaging}
  \bibinfo{volume}{40}, \bibinfo{pages}{2926--2938}.
\bibitem[{Wang et~al.(2021)Wang, Xia, Hu, Yan, Li, Wu, Huang, Gao, Metaxas and
  Zhang}]{wang2021few}
\bibinfo{author}{Wang, W.}, \bibinfo{author}{Xia, Q.}, \bibinfo{author}{Hu,
  Z.}, \bibinfo{author}{Yan, Z.}, \bibinfo{author}{Li, Z.},
  \bibinfo{author}{Wu, Y.}, \bibinfo{author}{Huang, N.}, \bibinfo{author}{Gao,
  Y.}, \bibinfo{author}{Metaxas, D.}, \bibinfo{author}{Zhang, S.},
  \bibinfo{year}{2021}.
\newblock \bibinfo{title}{Few-shot learning by a cascaded framework with
  shape-constrained pseudo label assessment for whole heart segmentation}.
\newblock \bibinfo{journal}{IEEE Transactions on Medical Imaging}
  \bibinfo{volume}{40}, \bibinfo{pages}{2629--2641}.
\bibitem[{Weston et~al.(2019)Weston, Korfiatis, Kline, Philbrick, Kostandy,
  Sakinis, Sugimoto, Takahashi and Erickson}]{weston2019automated}
\bibinfo{author}{Weston, A.D.}, \bibinfo{author}{Korfiatis, P.},
  \bibinfo{author}{Kline, T.L.}, \bibinfo{author}{Philbrick, K.A.},
  \bibinfo{author}{Kostandy, P.}, \bibinfo{author}{Sakinis, T.},
  \bibinfo{author}{Sugimoto, M.}, \bibinfo{author}{Takahashi, N.},
  \bibinfo{author}{Erickson, B.J.}, \bibinfo{year}{2019}.
\newblock \bibinfo{title}{Automated abdominal segmentation of ct scans for body
  composition analysis using deep learning}.
\newblock \bibinfo{journal}{Radiology} \bibinfo{volume}{290},
  \bibinfo{pages}{669--679}.
\bibitem[{Xia et~al.(2020)Xia, Yang, Yu, Liu, Cai, Yu, Zhu, Xu, Yuille and
  Roth}]{xia2020uncertainty}
\bibinfo{author}{Xia, Y.}, \bibinfo{author}{Yang, D.}, \bibinfo{author}{Yu,
  Z.}, \bibinfo{author}{Liu, F.}, \bibinfo{author}{Cai, J.},
  \bibinfo{author}{Yu, L.}, \bibinfo{author}{Zhu, Z.}, \bibinfo{author}{Xu,
  D.}, \bibinfo{author}{Yuille, A.}, \bibinfo{author}{Roth, H.},
  \bibinfo{year}{2020}.
\newblock \bibinfo{title}{Uncertainty-aware multi-view co-training for
  semi-supervised medical image segmentation and domain adaptation}.
\newblock \bibinfo{journal}{Medical Image Analysis} \bibinfo{volume}{65},
  \bibinfo{pages}{101766}.
\bibitem[{Yan et~al.(2022)Yan, Yang, Syer, Min, Punwani, Emberton, Barratt,
  Chiu and Hu}]{yan2022impact}
\bibinfo{author}{Yan, W.}, \bibinfo{author}{Yang, Q.}, \bibinfo{author}{Syer,
  T.}, \bibinfo{author}{Min, Z.}, \bibinfo{author}{Punwani, S.},
  \bibinfo{author}{Emberton, M.}, \bibinfo{author}{Barratt, D.},
  \bibinfo{author}{Chiu, B.}, \bibinfo{author}{Hu, Y.}, \bibinfo{year}{2022}.
\newblock \bibinfo{title}{The impact of using voxel-level segmentation metrics
  on evaluating multifocal prostate cancer localisation}, in:
  \bibinfo{booktitle}{Applications of Medical Artificial Intelligence: First
  International Workshop, AMAI 2022, Held in Conjunction with MICCAI 2022,
  Singapore, September 18, 2022, Proceedings},
  \bibinfo{organization}{Springer}. pp. \bibinfo{pages}{128--138}.
\bibitem[{Yu et~al.(2021)Yu, Dang, Tajbakhsh et~al.}]{yu2021location}
\bibinfo{author}{Yu, Q.}, \bibinfo{author}{Dang, K.},
  \bibinfo{author}{Tajbakhsh, N.}, et~al., \bibinfo{year}{2021}.
\newblock \bibinfo{title}{A location-sensitive local prototype network for
  few-shot medical image segmentation}, in: \bibinfo{booktitle}{2021 IEEE 18th
  International Symposium on Biomedical Imaging (ISBI)},
  \bibinfo{organization}{IEEE}. pp. \bibinfo{pages}{262--266}.
\bibitem[{Zhang et~al.(2019)Zhang, Lin, Liu, Guo, Wu and
  Yao}]{zhang2019pyramid}
\bibinfo{author}{Zhang, C.}, \bibinfo{author}{Lin, G.}, \bibinfo{author}{Liu,
  F.}, \bibinfo{author}{Guo, J.}, \bibinfo{author}{Wu, Q.},
  \bibinfo{author}{Yao, R.}, \bibinfo{year}{2019}.
\newblock \bibinfo{title}{Pyramid graph networks with connection attentions for
  region-based one-shot semantic segmentation}, in:
  \bibinfo{booktitle}{Proceedings of the IEEE/CVF International Conference on
  Computer Vision}.
\bibitem[{Zhao et~al.(2019)Zhao, Balakrishnan, Durand, Guttag and
  Dalca}]{zhao2019data}
\bibinfo{author}{Zhao, A.}, \bibinfo{author}{Balakrishnan, G.},
  \bibinfo{author}{Durand, F.}, \bibinfo{author}{Guttag, J.V.},
  \bibinfo{author}{Dalca, A.V.}, \bibinfo{year}{2019}.
\newblock \bibinfo{title}{Data augmentation using learned transformations for
  one-shot medical image segmentation}, in: \bibinfo{booktitle}{Proceedings of
  the IEEE/CVF conference on computer vision and pattern recognition}, pp.
  \bibinfo{pages}{8543--8553}.
\bibitem[{Zhou et~al.(2021)Zhou, Liu, Cao, Wei, Lu, Yu, Ma and
  Zheng}]{zhou2021generalized}
\bibinfo{author}{Zhou, H.Y.}, \bibinfo{author}{Liu, H.}, \bibinfo{author}{Cao,
  S.}, \bibinfo{author}{Wei, D.}, \bibinfo{author}{Lu, C.},
  \bibinfo{author}{Yu, Y.}, \bibinfo{author}{Ma, K.}, \bibinfo{author}{Zheng,
  Y.}, \bibinfo{year}{2021}.
\newblock \bibinfo{title}{Generalized organ segmentation by imitating one-shot
  reasoning using anatomical correlation}, in:
  \bibinfo{booktitle}{International Conference on Information Processing in
  Medical Imaging}, \bibinfo{organization}{Springer}. pp.
  \bibinfo{pages}{452--464}.

\end{thebibliography}

\end{document}